%% file: gravity.tex
\documentclass[alpha-refs]{wiley-article}
\usepackage{xspace}
\usepackage{graphicx}
\usepackage{balance}
\usepackage{amsmath}
\usepackage{amssymb}
\usepackage{subfig}
\usepackage{afterpage, float}
\usepackage{bm}
\usepackage{siunitx}

\newcommand{\eqtag}[1]%
    {}

\newcommand{\sens}{S}
\newcommand{\latent}{Q}
\newcommand{\vappress}{e}
\newcommand{\vp}{\vappress}
\newcommand{\refeqq}[1]{\mbox{(\ref{#1})}}
\newcommand{\reff}[1]{\mbox{figure \ref{#1}}}
\newcommand{\refs}[1]{\mbox{section \ref{#1}}}

\newcommand{\panel}[1]{\mbox{(#1)}}

\newcommand{\ttrop}{T_\textrm{trop}}

\papertype{Original Article}
\paperfield{Submitted to Quart.\ J.\ Roy.\ Met.\ Soc}

\input{mypreamble}
\newcommand{\Socrates}{\textsc{Socrates}\xspace}

\title{The Effects of Gravity on the Climate and Circulation of a Terrestrial Planet}

\author[1]{Stephen I. Thomson}
\author[1]{Geoffrey K. Vallis}

\affil[1]{University of Exeter, Exeter, EX4 4QF, UK}

\corraddress{Dr Stephen I. Thomson, College of Engineering, Mathematics and Physical Sciences, University of Exeter, Exeter, EX4 4QE, United Kingdom}

\corremail{stephen.i.thomson@gmail.com}

\fundinginfo{SIT is supported by the Natural Environment Research Council (grant number NE/M006123/1), and GKV acknowledges support from the Royal Society (Wolfson Foundation), the Leverhulme Trust and the Newton Fund.}

\runningauthor{Thomson and Vallis}

\begin{document} 
 \renewcommand{\topfraction}{0.95}
 \renewcommand{\textfraction}{0.05}
 
\maketitle

\begin{abstract}
  \linespread{1.0}\selectfont
The climate and circulation of a terrestrial planet are governed by, among other things, the distance to its host star, its size, rotation rate, obliquity, atmospheric composition and gravity. Here we explore the effects of the last of these, the Newtonian gravitational acceleration, on its atmosphere and climate. We first demonstrate that if the atmosphere obeys the hydrostatic primitive equations, which are a very good approximation for most terrestrial atmospheres, and if the radiative forcing is unaltered, changes in gravity have no effect at all on the circulation except for a vertical rescaling.  That is to say, the effects of gravity may be completely scaled away and the circulation is unaltered. However, if the atmosphere contains a dilute condensible that is radiatively active, such as water or methane, then an increase in gravity will generally lead to a cooling of the planet because the total path length of the condensible will be reduced as gravity increases, leading to a reduction in the greenhouse effect. Furthermore, the specific humidity will decrease, leading to  changes in the moist adiabatic lapse rate, in the equator-to-pole heat transport, and in the surface energy balance because of changes in the sensible and latent fluxes. These effects are all demonstrated both by theoretical arguments and by numerical simulations with moist and dry general circulation models.
\end{abstract}

~  \\

\section{Introduction}

The climate of a terrestrial planet depends on an almost uncountable number of factors, including the distance to its host star, the nature of that host star, the size and rotation rate of the planet,  the atmospheric composition and many other factors. The variety of planetary climates is large, and there is and can be no single theory of planetary climate, nor is there a planetary analogue of the Hertzsprung--Russell diagram showing how the luminosity of stars varies with their effective temperature. However, this is not to say that we cannot apply general physical principles to atmospheric circulation and planetary climate. Thus, for example, \citet{Read11} and \citet{Wang_etal18} describe how various nondimensional parameters describe the general circulation of a large class of planetary atmospheres, \citet{Kaspi_Showman15} illustrate how the planetary circulation patterns vary over a wide range of orbital parameters,  and \citet{Pierrehumbert10} applies building blocks based on elementary physical principles to construct a plentiful panoply of planetary climates. 

As regards planetary circulation, among the most studied parameters are the planetary radius and rotation rate; these  combine to give the external Rossby number that is one of the single most influential parameters on planetary circulation. Atmospheric composition obviously plays a key radiative role in determining the surface temperature, especially if the composition gives rise to a greenhouse effect, or an anti-greenhouse effect, and the changing composition of Earth's atmosphere is obviously of current interest. 

Less well studied is the effect of gravity, here meaning the Newtonian gravitational acceleration as measured at the planetary surface. One expects that a planet with a higher gravity than another, but otherwise the same, would have a thinner (meaning less extended) atmosphere with a higher surface density, but the effects on the circulation and temperature are less clear. The matter was partially investigated by \citet{Kilic_etal17} and \citet{Kaspi_Showman15}, but their model set-ups were very different and their results  too incompatible to  compare, with the former fixing their surface temperatures independent of gravity, and the latter using a simplified GCM without many Earth-like effects, such as the radiative effect of water vapour.  In this paper we revisit the issue, looking at it both as problem in geophysical fluid dynamics and a problem in planetary climate. 

We first, in \secref{sec:dry}, examine how the adiabatic equations of motion, both the primitive equations and the full Navier--Stokes equations, scale with gravity.  We find that in the primitive equations the effects of gravity can be completely scaled out of the problem and that, if the diabatic forcing is sufficiently simple,  the circulation is unaltered. This invariance is broken both by non-hydrostatic effects and by having a non-shallow atmosphere, but in many planetary atmospheres these effects will be small, although not always negligible \citep{Mayne_etal18}. In Section \ref{sec:moisture_content}
we describe how changes in gravity lead to non-negligible changes in changes in moisture content.  We then explore the effects of these changes using some idealized numerical simulations: first, in Section \ref{sec:radiative}, we describe the radiative effects of those changes,  and then in Sections \ref{sec:specific_humidity_effects}--\ref{sec:tropics} we explore the dynamical effects of the changes in specific humidity,  In \refs{sec:socrates} we look at the role of gravity with a more complete model, and \refs{sec:conclusions} we provide our conclusions.

\section{Invariance of the Equations of Motion} \label{sec:dry}
The momentum equations in the primitive equations on the sphere may be written, in standard notation, as \citep{Vallis19}
\begin{gather}
\label{prim.1}
\DD u - 2 \Omega v \sin\vrt  + {u v \tan \vrt \over a } 
  = - {1 \over \rho a \cos \vrt} \pp p \lambda, \\
\DD v +  2 \Omega u \sin \vrt + {u^2 \tan \vrt \over a } 
= - {1 \over \rho a }\pp p \vrt, \\
\pp p z  = - \rho g.
\end{gather}
The mass continuity and adiabatic thermodynamic equations are, respectively,
\begin{equation} 
    \DD \rho + \rho \div \vb = 0 ,
    \label {prim.32}
\end{equation}
and
\begin{equation} 
    c_v \DD T + \frac p \rho \div \vb = 0 , \qquad \text{or} \quad \DD \theta = 0 .
    \label {prim.3}
\end{equation}
where $\vb$ is the three-dimensional velocity, $\lambda$ is longitude, $\vrt$ is latitude, $T$ and $\theta$ are temperature and  potential temperature,  and the other notation is quite standard.
These equations remain invariant under the following transformation:
\begin{equation}
	\label{prim.4} 
    \begin{gathered}
   	g \to \alpha g,  \qquad  p \to \alpha p , \qquad \rho \to \alpha \rho, \qquad  (T, \theta) \to (T, \theta),\\
    t \to t,  \qquad (x, y) \to (x, y), \qquad    z \to z/\alpha, \\
    \qquad (u, v) \to (u, v), \qquad w \to w/\alpha.
     \end{gathered}  
\end{equation}
If we substitute \eqref{prim.4} into \eqref{prim.1}--\eqref{prim.3} then all the factors of $\alpha$ cancel and the equations are unchanged, as was noted by \citet{Vallis19}.  Given the invariance of the unforced equations themselves, it is a simple matter to confirm that all quantities of dynamical interest, such as the deformation radius, $L_d \equiv NH/f$ and the Eady growth rate, $0.31 U/L_d$, remain invariant. 

This invariance does not hold in the full Navier--Stokes equations on the sphere.   The full momentum equations are
\begin{gather}
\label{prim.5}
\DD u - \left( 2 \Omega + {u \over r \cos \vrt} \right) (v \sin \vrt - w \cos \vrt)
  = - {1 \over \rho r \cos \vrt} \pp p \lambda, \\
\DD v + { w v \over r} + \left( 2 \Omega + {u \over r \cos \vrt} \right) u \sin \vrt
= - {1 \over \rho r }\pp p \vrt, \\
\DD w - {u^2+v^2 \over r} - 2 \Omega  u \cos \vrt
= - {1 \over \rho} \pp p r - g.
\end{gather}
The additional metric terms in the horizontal momentum equations, for example $uw/r$, and the vertical acceleration term $\DDD w$ in the vertical equations, are not invariant with respect to the transformation. The importance of these terms depends on the ratio of the thickness of the atmosphere to the radius of the planet and this is quite small in most terrestrial atmospheres -- in the Solar System Titan perhaps comes closest to refuting this statement: Titan has a radius is 2576\km, a scale height of about 20\km, a tropopause at about 40\km and a stratopause at about 300\km, still only 12\% of the planetary radius.  Non-hydrostatic motion within an atmosphere also violates the invariance, as is implicit in the `hypo-hydrostatic' rescaling of \citet{Garner_etal07}.  Finally, terms on the right-hand side of the thermodynamic equation might also violate the invariance, as we consider later.

\subsection{Simulations with primitive equations and Newtonian relaxation}

\begin{figure*}
\centering
\subfloat[]{
\includegraphics[width=0.5\textwidth]{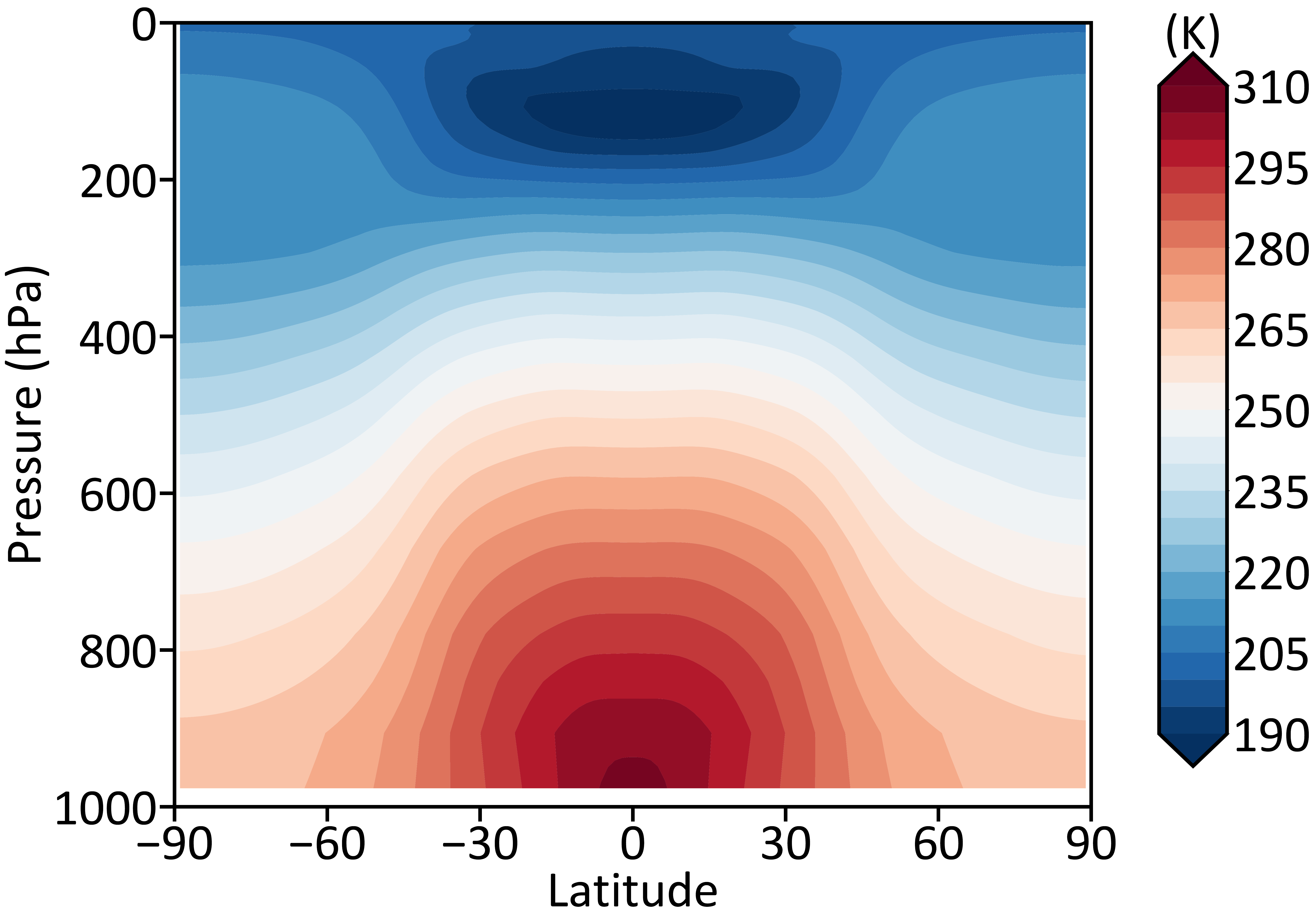}
}
\subfloat[]{
  \includegraphics[width=0.5\textwidth]{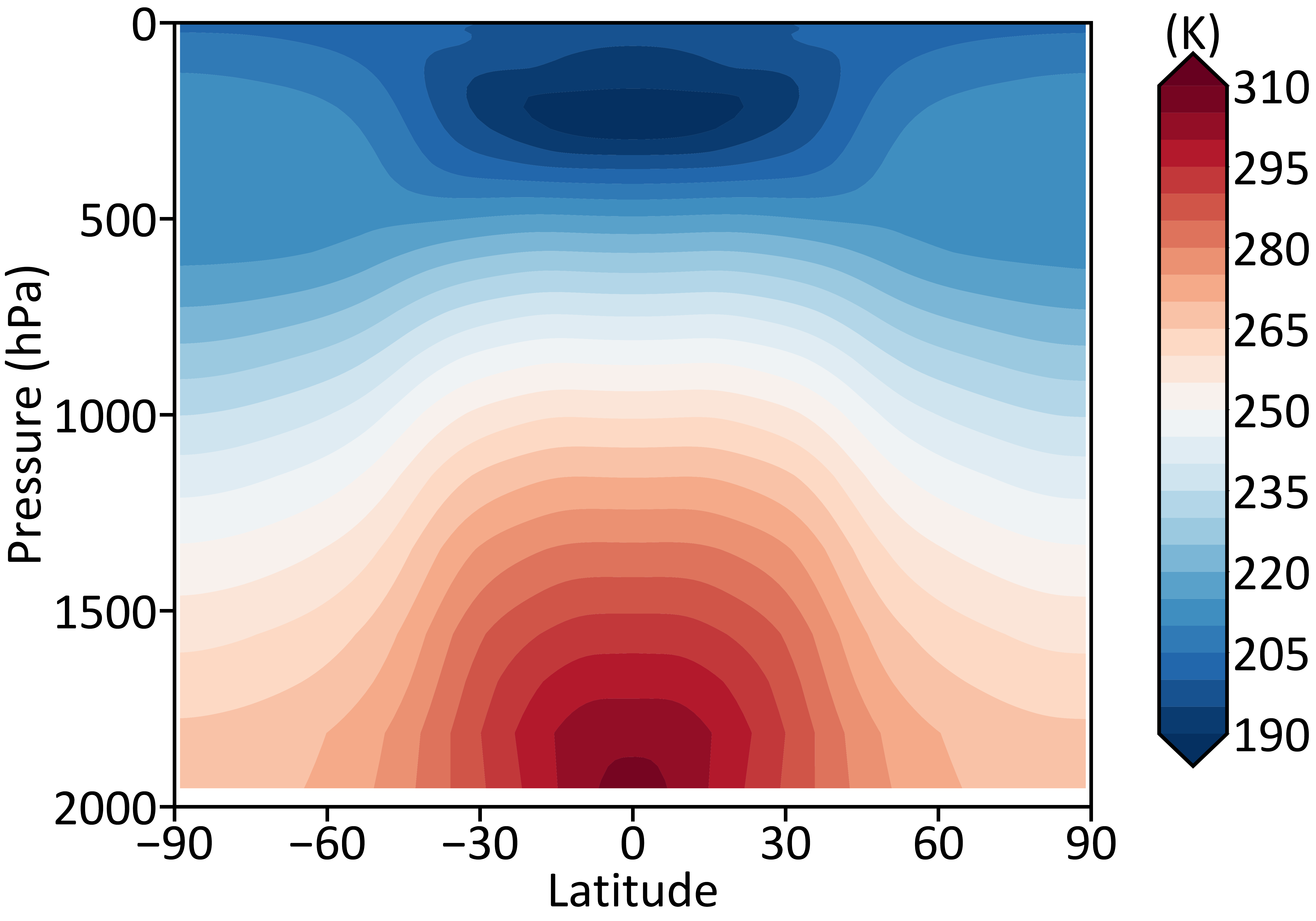}
}
\caption{\label{fig:hs1} (a) The zonal-mean temperature as a function of pressure and latitude in a Held-Suarez run with normal Earth gravity. (b) The same but with twice Earth gravity.}
\end{figure*}

To demonstrate how the invariance manifests itself in practice we perform simulations with a dry dynamical core obeying the primitive equations, using the Isca framework \citep{Vallis_etal18}. The forcing is that of \citet{Held_Suarez94}, which is a Newtonian relaxation back to a specified temperature that is a function of pressure and latitude. We perform an integration with the standard value of gravity  ($9.8\m\s^{-2}$) and one with double that value, keeping the total mass of the atmosphere constant in the two integrations.  \Figref{fig:hs1} shows the temperature field in the two integrations and, as is evident, they are identical (in their early stages they are bit-wise identical, but numerical artifacts mean that their final state is not). The velocity and pressure fields (not shown) are also identical. If the fields were plotted in height co-ordinates then the case with doubled gravity would appear as flatter (with $z\to z/2$), but this has no dynamical effect in the primitive equations. 

The invariance discussed above does not necessarily hold if we add more realistic forcing to the thermodynamic equation, and in particular if the radiative forcing is sensitive to gravity. Although the effects can be quite subtle, those due changes in the moisture content, or any other radiatively active condensible, are more clear. We investigate some of these effects in the following sections.

\section{Change in Moisture Content} \label{sec:moisture_content}

There are two distinct changes that a condensible may bring about as gravity changes, one due to its radiative properties if it is a greenhouse gas and the other due to the release of latent heat when it condenses. For specificity we deal with water vapour, and assume that the condensible is dilute (meaning the fraction of the condensible is small).   Both of the effects arise because the fraction of condensible, relative to the rest of the atmosphere, will diminish if gravity increases, as we discuss below. The overall temperature of the planet's surface  will then diminish as gravity increases (since water vapour is a potent greenhouse gas), and the dynamical effects of condensation (for example, in setting the saturated adiabatic lapse rate) will also diminish.

\subsection{Total water vapour content} 
\label{sec:radiative_effects}
 
The total water vapour content of a column of atmosphere, $W$, is given by
\begin{equation}
	\label{rad.1} 
   	W = \int_0^\infty {\vappress \over R_vT}  \dz.
\end{equation}
where $\vappress$ is the vapour pressure of the water vapour, $R_v$ is its specific gas constant and $T$ is the temperature.  Suppose that we increase the gravity of a planet by a factor $\alpha$, without initially changing the temperature.  The vapour pressure is a strong function of temperature, and at fixed relative humidity is only a function of temperature (since saturation vapour pressure is a function only of temperature in an ideal gas).  
If gravity increases by a factor $\alpha$ then the lowest order effect is for temperature to fall with height by a factor $\alpha$ more rapidly than before, following the scaling of \secref{sec:dry}. That is, at any given height $z$ the value of $\vappress$ will be lower that before, and the total water content of the atmosphere will fall roughly by a factor $\alpha$.  Changes in relative humidity can quantitatively alter this conclusion but unless relative humidity also changes by a factor $\alpha$, which is in most circumstances very unlikely, that change will be small.

To see the above argument another way, we transform \eqref{rad.1} into an integral over pressure and, using the hydrostatic and ideal gas relations, obtain
\begin{equation}
	\label{rad.2} 
   	W = -\frac{1}{g}\int_0^\infty \vappress { R_d \over R_v }  \d \log (p/p_s).
\end{equation}
where $R_d$ is the specific gas constant for air, $p_s$ is the surface pressure, and  the limits of the integral are the same even as $g$ and $p_s$ change. Now, as noted above, $\vappress$ is determined largely by temperature,  and the value of temperature at any given value of $\log (p/p_s)$ is, to lowest order, unaltered (as in \figref{fig:hs1}).  Thus, the integrand is unaltered by the transformation, but the factor of $1/g$ outside the integral indicates that the total water content will scale by a factor of $1/\alpha$.  Of course, once the water content changes the temperature changes  because of radiative effects, which causes the water content to change again, so the effect is not a simple one. Nonetheless,  the most basic effect that can be expected is that if gravity increases water vapour content will fall.  Since water vapour is a potent greenhouse gas, temperature will also fall. 

\subsection{Specific humidity} \label{sec:specific_humidity}

In addition to the total water vapour content, the specific humidity, $q$, will also fall as gravity increases and this can have an important dynamical effect.  The specific humidity is defined as the ratio of the mass of water vapour to the total mass of air and in terms of pressures it may be written
\begin{equation}
	\label{spec.1} 
   	q = \frac{\eps \vp} {p - \vp(1 - \eps)} \approx \frac{\eps \vp} p,
\end{equation}
where $\eps$ is the ratio of the molar mass of water vapour to that of dry air and the approximation giving $\eps \vp/p$ holds for a dilute atmosphere. Since pressure scales with $\alpha$ but $\vp$ does not (it does not depend on $\alpha$ at lowest order) we expect that the specific humidity will fall as gravity increases, scaling roughly as $1/\alpha$.  (Note that $vp = \mathcal H \vp_s$ where $H$ is relative humidity and $\vp_s$ is the saturation vapour pressure, a function only of temperature. As with the argument for total water vapour content, unless $\mathcal H$ changes, $\vp$ will not change as $\alpha$ changes.)

The consequence of this is that the hydrology cycle will weaken as gravity weakens, essentially because the condensation will have a smaller effect on a denser atmosphere. That is, if there is a change in specific humidity of $\Delta q$ then the temperature change is given by 
\begin{equation}
	\label{spec.2} 
   	\cp \Delta T = - L \Delta q,
\end{equation}
so that $\Delta T$ is smaller as $q$ falls.

Thus, in summary, an increase in gravity has two somewhat distinct effects on the condensible (with the opposite effect for a decrease in gravity). First, the total amount of condensible decreases, roughly in proportion to the increase in gravity, because of the reduced scale height of the temperature field. One effect of this is to reduce the greenhouse effect of the condensible and so make the atmosphere cooler. Second, the specific humidity falls in a dilute atmosphere, not primarily because of the cooling of the atmosphere but because of the increase in total pressure of the atmosphere and the approximate constancy of the vapour pressure, $\vappress$, as given by \eqref{spec.1}. This effect will be further amplified by the cooling of the planet because of the reduced greenhouse effect, but does not depend upon it. The main consequences of this are that the hydrology cycle will weaken and, concomitantly, the magnitude of the saturated adiabatic lapse rate will increase as it approaches the dry adiabatic lapse rate. 

 In the sections that follow we first explore and quantify the radiative effect, and then the dynamical effects of the weaker hydrology cycle, in both cases we using idealized radiative transfer schemes to isolate the effects. In Section \ref{sec:socrates} we use a more accurate radiative transfer scheme to see how the effects work together.

\section{Radiative Effects}  \label{sec:radiative}

As noted above water vapour is a greenhouse gas so that increasing gravity will lead to a cooling of the atmosphere.  We illustrate this effect by a set of integrations with a moist general circulation model, again using the Isca framework. In all of the following experiments the gravitational acceleration is changed, and the model's mean surface pressure is prescribed to change like $\alpha$, representing a constant atmospheric mass between experiments.  We configure Isca to use a grey radiative transfer with an optical depth prescription that depends on the atmospheric specific humidity, $q$. We follow \citet{Byrne_OGorman13}  except that we change the parameter $a=0.1627$ so that the time-averaged surface temperatures are similar to that achieved with a complex radiative transfer code with a surface albedo of $0.3$, as discussed in \citet{Vallis_etal18}. In other respects the model is similar to that described in \citet{Frierson_etal06}, except that virtual temperature effects are included.  For simplicity we omit the seasonal cycle using instead a time-constant insolation profile  which well approximates annual mean insolation on Earth. 

The zonal-mean surface temperature profiles under different gravitational accelerations are shown in \reff{fig:bog1}\panel{a}. Additionally, global-average values of $W$ are plotted against $1/\alpha$ in panel \panel{b}, and the response of the zonal-mean atmospheric temperature to doubling gravity is shown in panel \panel{c}. The latter is presented in so-called `sigma' coordinates, where $\sigma = p/p_{\textrm{surf}}$, allowing the difference between Earth gravity and twice-Earth gravity to be presented on one plot.  In panel \panel{b}, the increase in $W$ with increasing $1/\alpha$ is consistent with expectations for the decrease in $W$ with increasing gravity. The slope is, however, different from a simple $1/\alpha$ dependence, owing to the increase in gravity and the concomitant decrease in temperature, both of which act to decrease $W$ for increasing $\alpha$. Alongside a decrease in $W$ with increasing $\alpha$, the associated decrease in long-wave optical depth and subsequent surface cooling is evident in panel \panel{a}.

\afterpage{
\begin{figure}[H]
\centering
\includegraphics[width=0.5\textwidth]{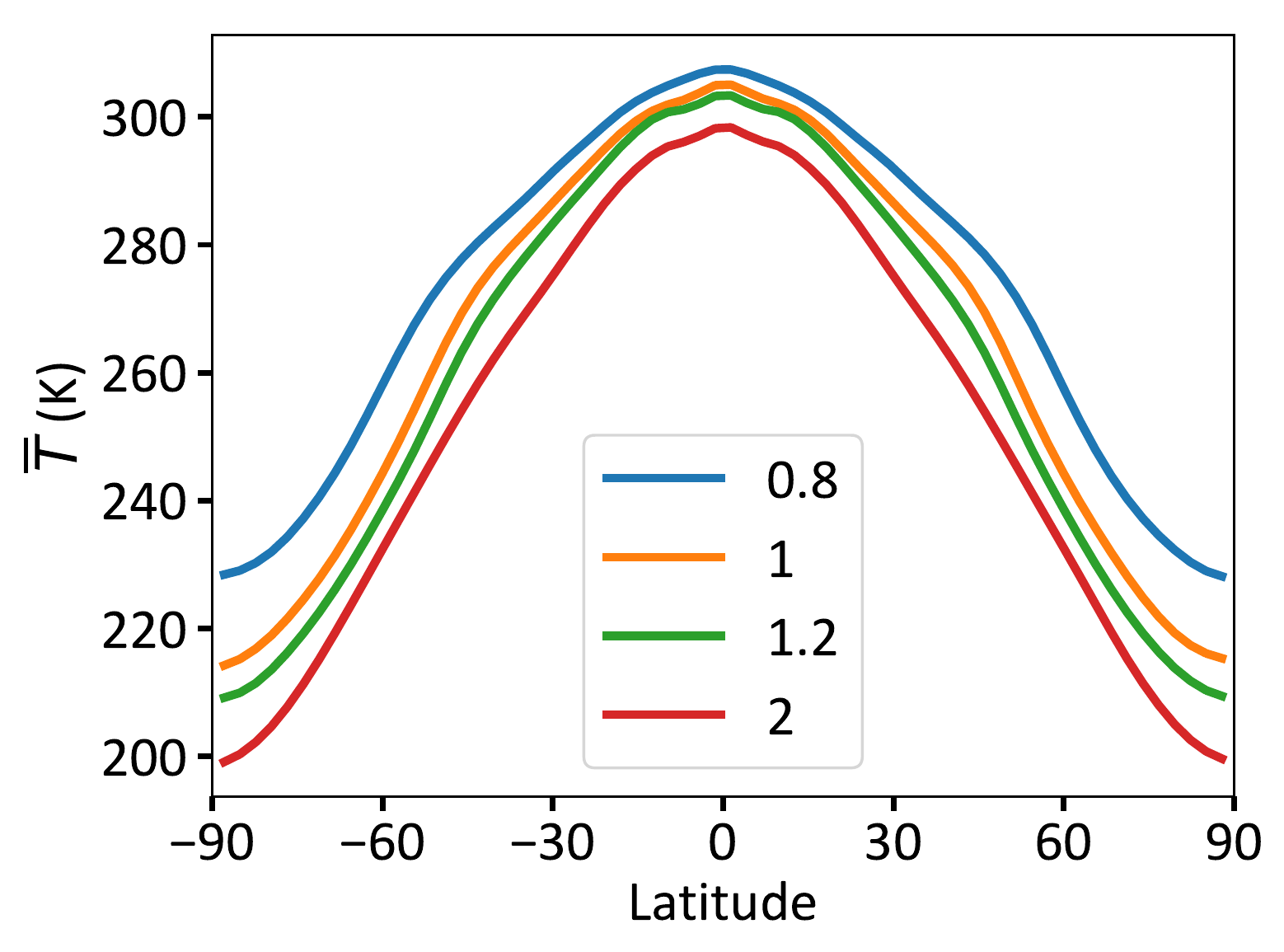}%
  \includegraphics[width=0.5\textwidth]{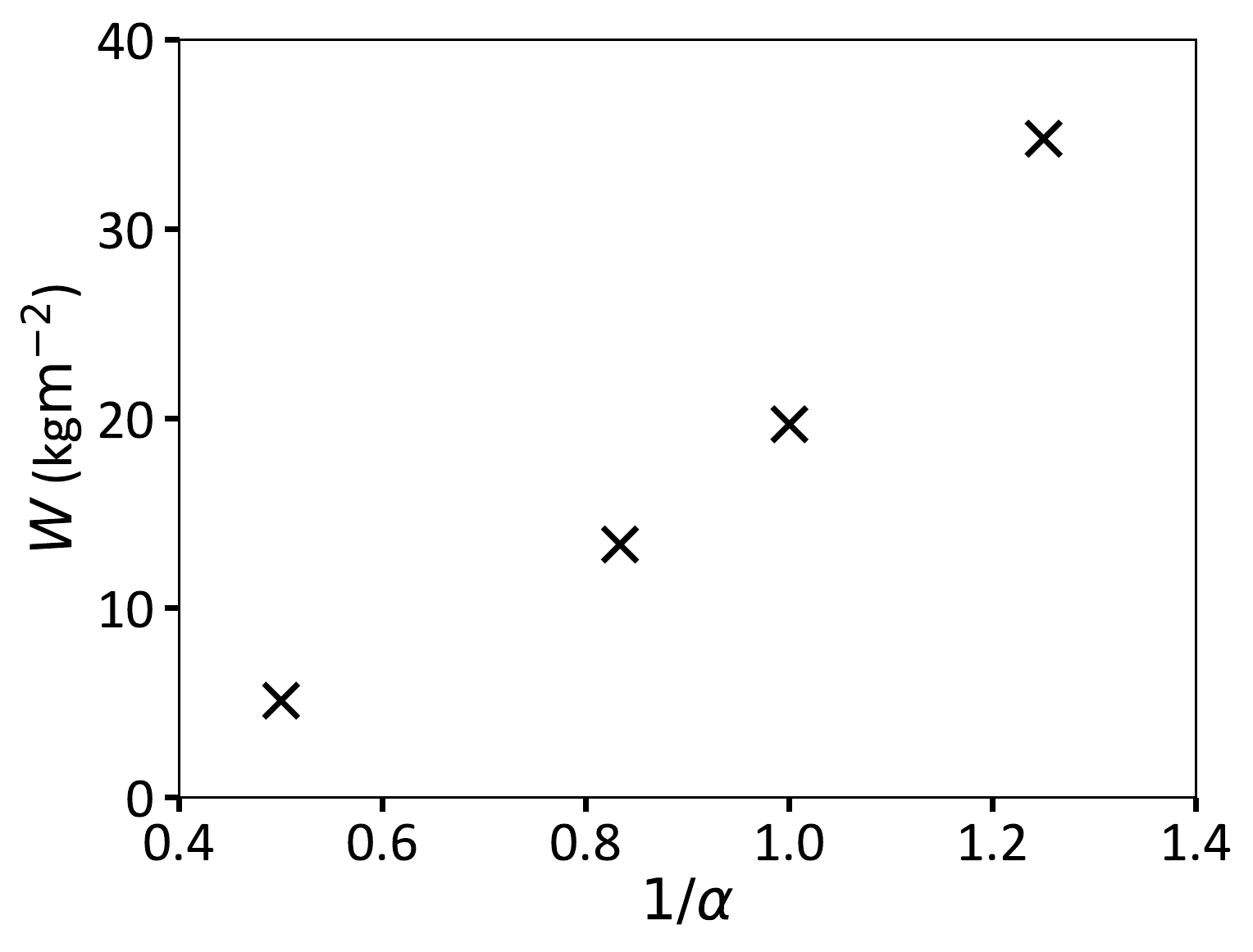}
\\
  \includegraphics[width=0.5\textwidth]{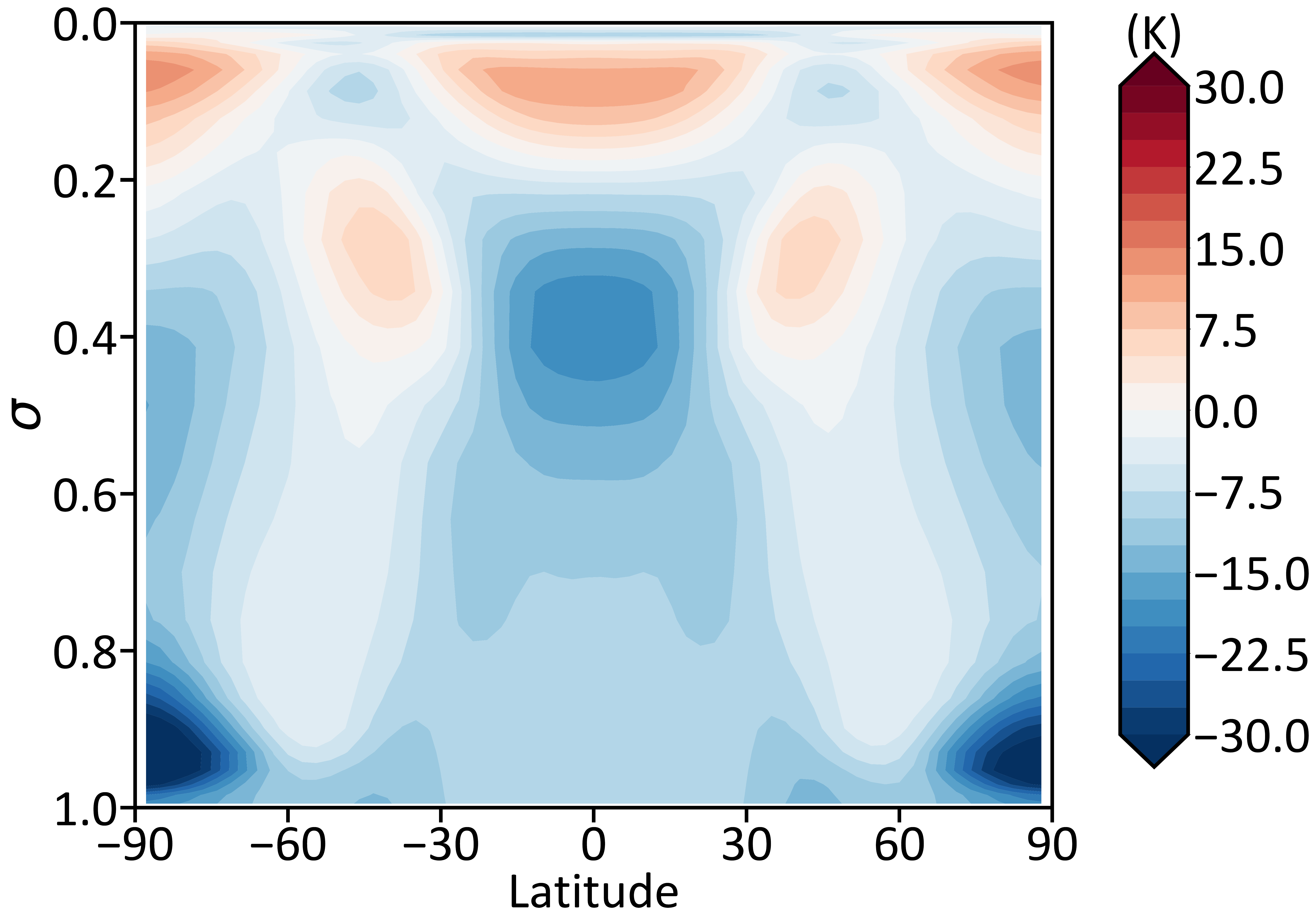}
\caption{\label{fig:bog1} (a) The zonal-mean surface temperature against latitude in aquaplanet run with optical depths like \citet{Byrne_OGorman13}. (b) The time and area-averaged value of $W$ plotted against $1/\alpha$. (c) The zonal-mean atmospheric temperatures in a twice-Earth-gravity run minus the same in a normal-Earth-gravity run, plotted as a function in $\sigma$ coordinates.}
\end{figure}
}

\section{Specific Humidity Effects} \label{sec:specific_humidity_effects}

In addition to the overall cooling, the structure of the cooling has a distinctive pattern, as can be seen in panel \panel{c} of \reff{fig:bog1}.  Two effects are particularly noticeable: an enhanced cooling in both the tropical upper troposphere and near the surface at high latitudes (a `polar amplification').  In this section we determine the  mechanisms determining this structure, with more details in sections following.
 
These effects are essentially the inverse of a global-warming response, illustrated for example in cf. figure 6 of \citet{Vallis_etal15}, and the mechanisms are similar (but inverted), and are due to the changes in specific humidity.   To isolate the effect we perform the same set of experiments as those described above using a radiative scheme with a fixed optical depth (and so one that does not depend on water-vapour amount), as in \citet{Frierson_etal06}. We thereby eliminate the overall global cooling effect.

The profiles of zonal-mean surface temperature in these runs are shown in \reff{fig:frierson1}\panel{a}. It is clear from comparison between this figure and \reff{fig:bog1}\panel{a} that removing the water-vapour--optical-depth feedback has altered the response to changing gravity considerably. Without the long-wave optical depth feedback, the twice gravity profile is now warmer in the tropics and colder at the poles than its Earth-gravity equivalent, unlike the response seen with the feedback. The increase in total column water vapour, $W$, with increasing $1/\alpha$ shown in panel \panel{b} is also present, but scaling is closer to $1/\alpha$ than in figure \ref{fig:bog1}\panel{b}, owing to the lack of global cooling in the newer experiments. The atmospheric temperature response is also different, as seen by comparing panel \panel{c} in \reff{fig:bog1} with panel \panel{c} in \reff{fig:frierson1} -- the enhanced low level cooling over the poles is not as conspicuous, and this is because this cooling relies in part on a direct radiative effect not present in the runs with fixed optical depth.  

However, the enhanced upper level cooling in the tropics is still present, and this is due to changes in saturated adiabatic lapse rate. As $q$ diminishes then the saturated adiabatic lapse rate increases in magnitude, so that in the tropics upper levels cool preferentially, as can be seen in panel (c) in both \figref{fig:bog1} and \figref{fig:frierson1}. Changes in the $q$ profiles in the latter experiments are shown in panel \panel{d} of \figref{fig:frierson1}. We explore the lapse-rate effect further in \secref{sec:tropics}.

Interestingly, it is found that the poleward moist-static-energy flux, or `heat transport' (not shown) changes very little between the experiments with varying gravity.   The moist-static energy flux is given by
\begin{equation}
    \overline{v MSE} = \int v C_p T + v g z + v L_v q \dx\dz
\end{equation}    
where $MSE$ is the moist static energy and $v$ is the meridional velocity. Despite the consistency in the overall transport, the balance of terms in this equation does change. Specifically increasing gravity decreases $q$, increases $v$ and also changes the temperature structure, thereby affecting all the terms. The lack of change in overall transport is consistent with results found in an idealised GCM by \citet{Frierson_etal07} and references therein.   However, unchanged overall transport when changing parameters is certainly not always the case \citep[e.g.][]{Schneider_etal10}.   

The changes in temperature structure, then, are not primarily caused by changes in overall heat transport. Rather, further investigation indicates that these changes have two main causes, namely changes in the fluxes from the surface to atmosphere, and (as previously noted) changes in the tropical lapse rates. We now discuss each of these in turn.

\afterpage{
\begin{figure}[H]
\centering
\includegraphics[width=0.45\textwidth]{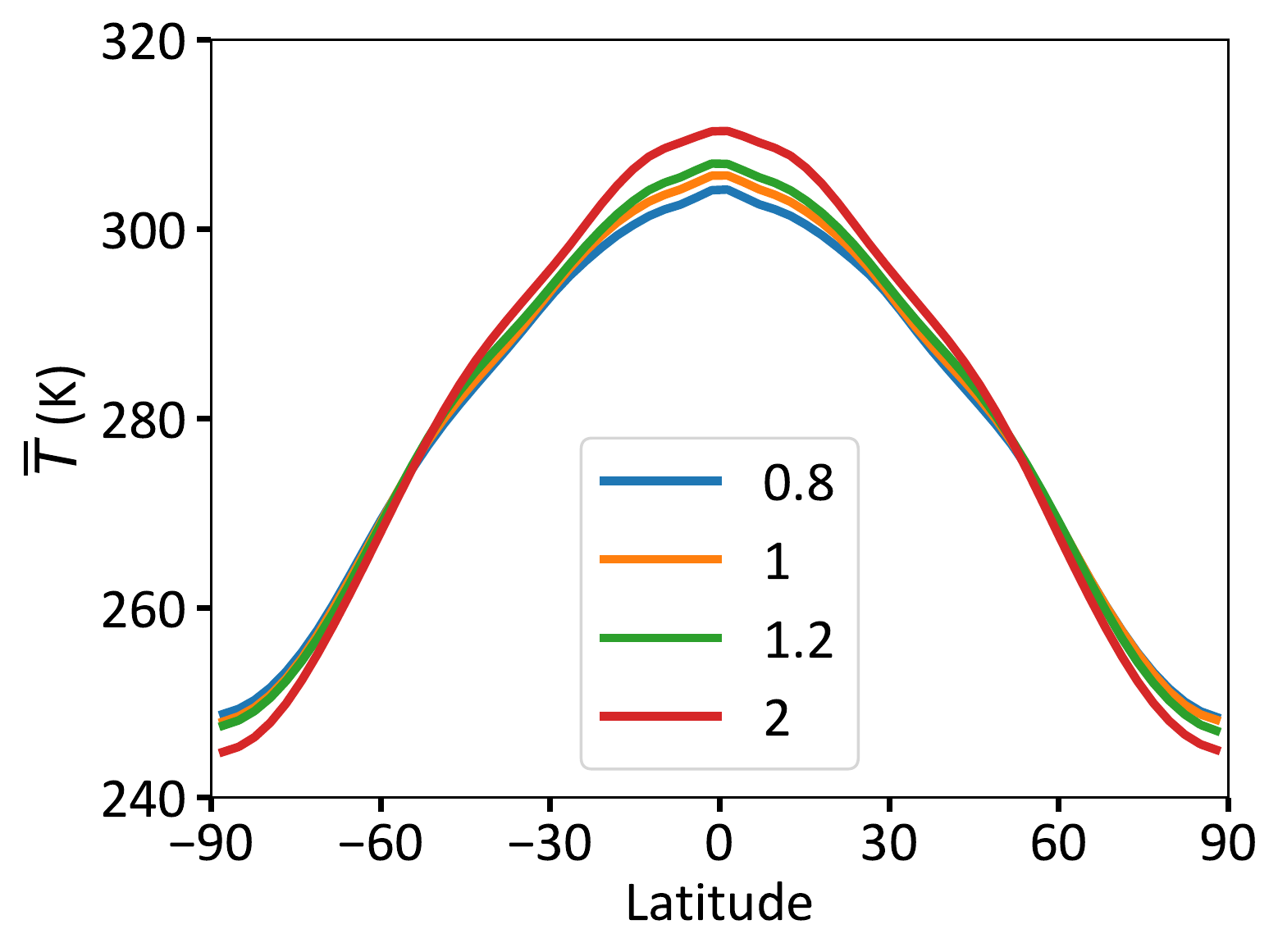}
  \includegraphics[width=0.45\textwidth]{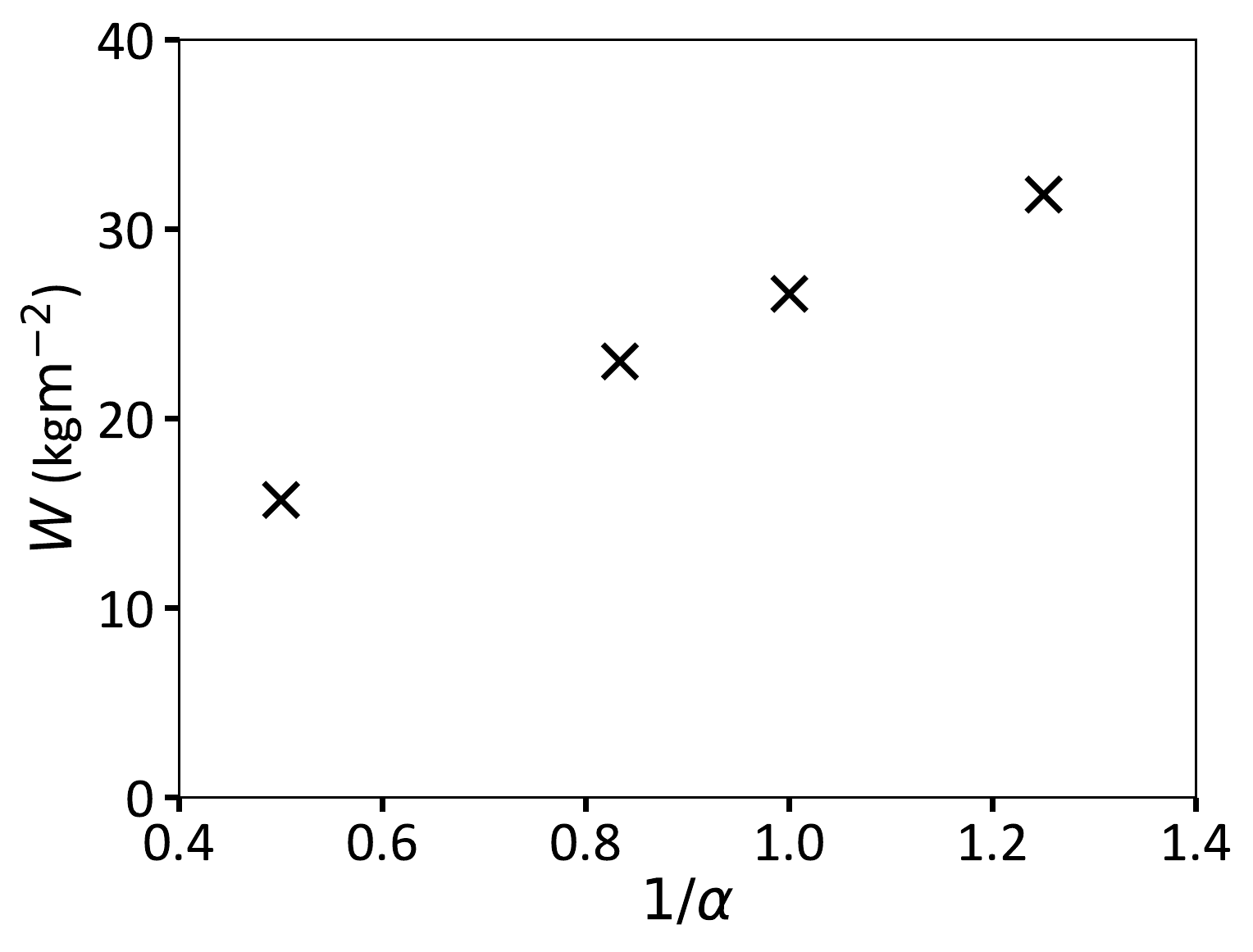} \\
  \includegraphics[width=0.45\textwidth]{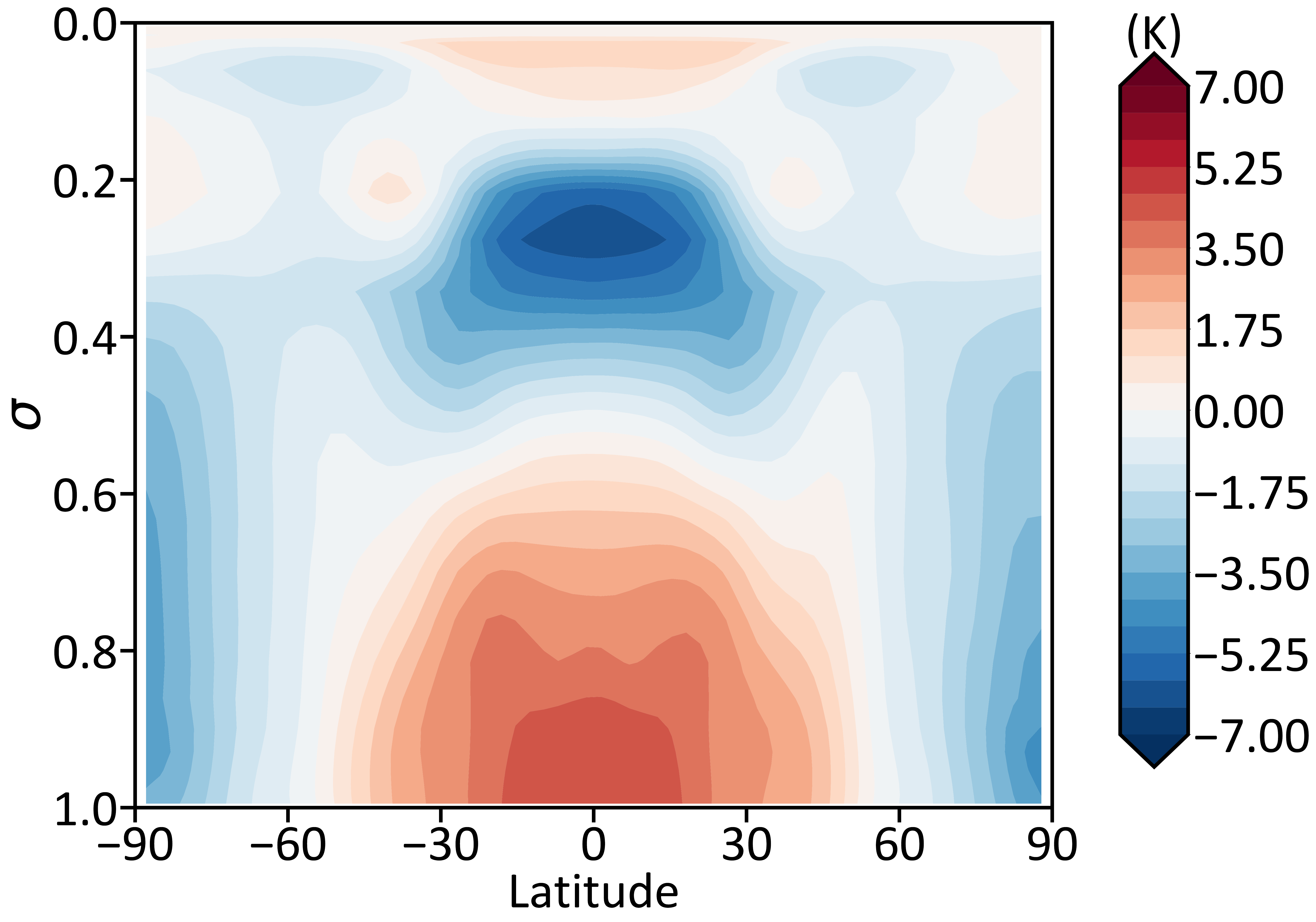}
  \includegraphics[width=0.45\textwidth]{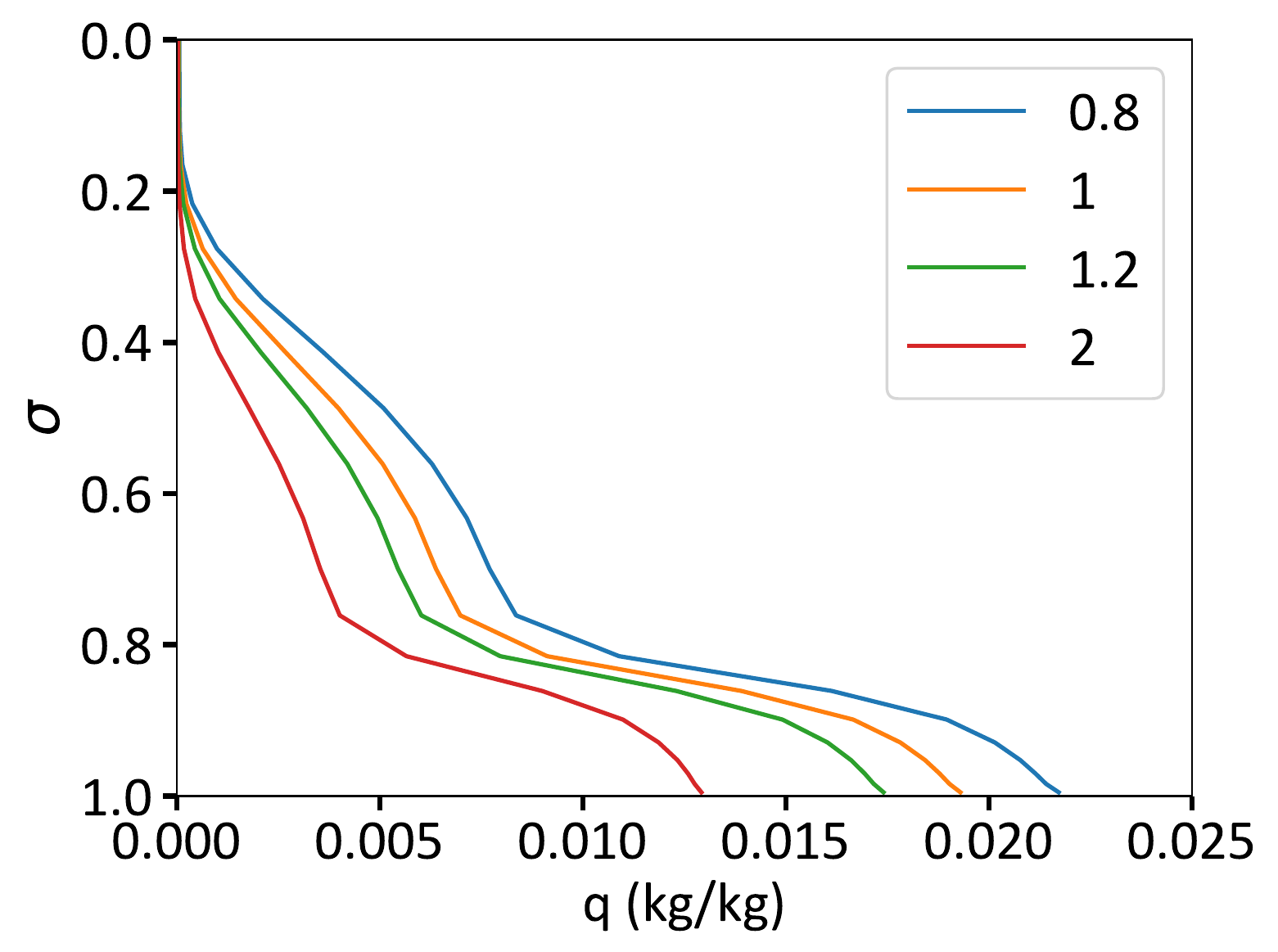}
\caption{\label{fig:frierson1} Panel \panel{a} The zonal-mean surface temperature against latitude in aquaplanet run with optical depths like \citet{Frierson_etal06}. Panel \panel{b} The time and area-averaged value of $W$ plotted against $1/\alpha$.  \panel{c} The zonal-mean atmospheric temperatures in a twice-Earth-gravity run minus the same in a normal-Earth-gravity run, plotted as a function in $\sigma$ coordinates. \panel{d} Vertical profiles of $q$ averaged between $10^\circ$S and $10^\circ$N.}
\end{figure}
}

\newcommand{\Tbar}{\overline T}
\section{Surface-Flux Effects}
\label{sec:surface}

\subsection{A vertical-flux-based argument}

Consider now the effects of heat, momentum and moisture exchange between the surface and the lower atmosphere. The effects on the atmosphere can be written as the vertical gradient of upward eddy fluxes of the relevant quantity:
\begin{subequations}
\begin{align}
  \frac{\partial T }{\partial t} & = ... - \frac{\partial}{\partial z} \left( \overline{T'w'} \right) \\
  \frac{\partial u  }{\partial t} & = ... - \frac{\partial}{\partial z} \left( \overline{u'w'} \right)  \\
  \frac{\partial q }{\partial t} & = ... - \frac{\partial}{\partial z} \left( \overline{q'w'} \right) ,
\end{align} 
\label{eq:vertical_flux_boundary_fluxes}
\end{subequations} 
with the overbars representing a mean over some area, and the primes being a departure from that mean -- these are subgridscale quantities in a GCM.    If we consider how each of these flux terms scale with a change in gravity, an obvious difference between them is that $T$ and $u$ do not scale simply with $\alpha$, but $q$ scales like $1/\alpha$, as discussed in \refs{sec:specific_humidity}.  In addition, the fluxes themselves will vary with gravity, as we now show.

If we were to scale the various terms in \eqref{eq:vertical_flux_boundary_fluxes} using \eqref{prim.4}, then all factors of $\alpha$ cancel and it might appear that the surface fluxes are unaltered. However, this scaling is unwarranted because the fluxes are non-hydrostatic.   We would expect $z$ to still scale like $1/\alpha$, as descried in section \refs{sec:dry}, but in the boundary-layer the turbulence is essentially isotropic,  meaning that $w'$ scales like $u'$, which does \textit{not} scale with $\alpha$. This suggests that the tendencies of the vertical flux terms will scale like $\alpha$ for $u$ and $T$. The tendencies from the $q$ term do not scale with $\alpha$, but if we account for the factor of $1/\alpha$ on the LHS, then the effect of the vertical flux on the scaled $q$ also scales like $\alpha$. 

The same conclusions can be drawn if we formulate the boundary-layer fluxes as diffusion terms of the form
\begin{equation}
    \frac{\partial T}{\partial t} = ...  \frac{\partial}{\partial z} \left( \kappa \frac{\partial T}{\partial z} \right) 
    \label{eq:vertical_flux_boundary_fluxes2}
\end{equation}
and similarly for $u$ and $q$. We expect the eddy diffusivity,  $\kappa$, to scale like an eddy velocity multiplied by a vertical length-scale, $l' w' $. Now, $w'$  does not scale with $\alpha$ (since $w' \sim u'$),  whereas $l'$ scales like $1/\alpha$, and so $\kappa$ itself scales like $1/\alpha$. The right-hand side of \eqref{eq:vertical_flux_boundary_fluxes2} then scales like $\alpha$, as before.

\subsection{Surface flux implementation in Isca}

The surface fluxes in Isca, as in most GCMs, are parameterized with bulk-aerodynamic laws, but these obey the same scalings as above as we will show.  First consider the simplest case of a neutrally-stable boundary-layer over a smooth surface. Here, it is common to take $\kappa = K_\text{vk} u_* z$ \citep[e.g.,][equation (5.15)]{Kraus72} where $K_\text{vk} \approx 0.4$ is the Von Karman constant and $u_*$ is the turbulent velocity, which does not scale with $\alpha$. Thus,  $\kappa \sim 1/\alpha$, and the effects of the boundary-layer flux convergence scales like $\alpha$. 

\afterpage{
\begin{figure}[H]
\centering
\includegraphics[width=0.49\textwidth]{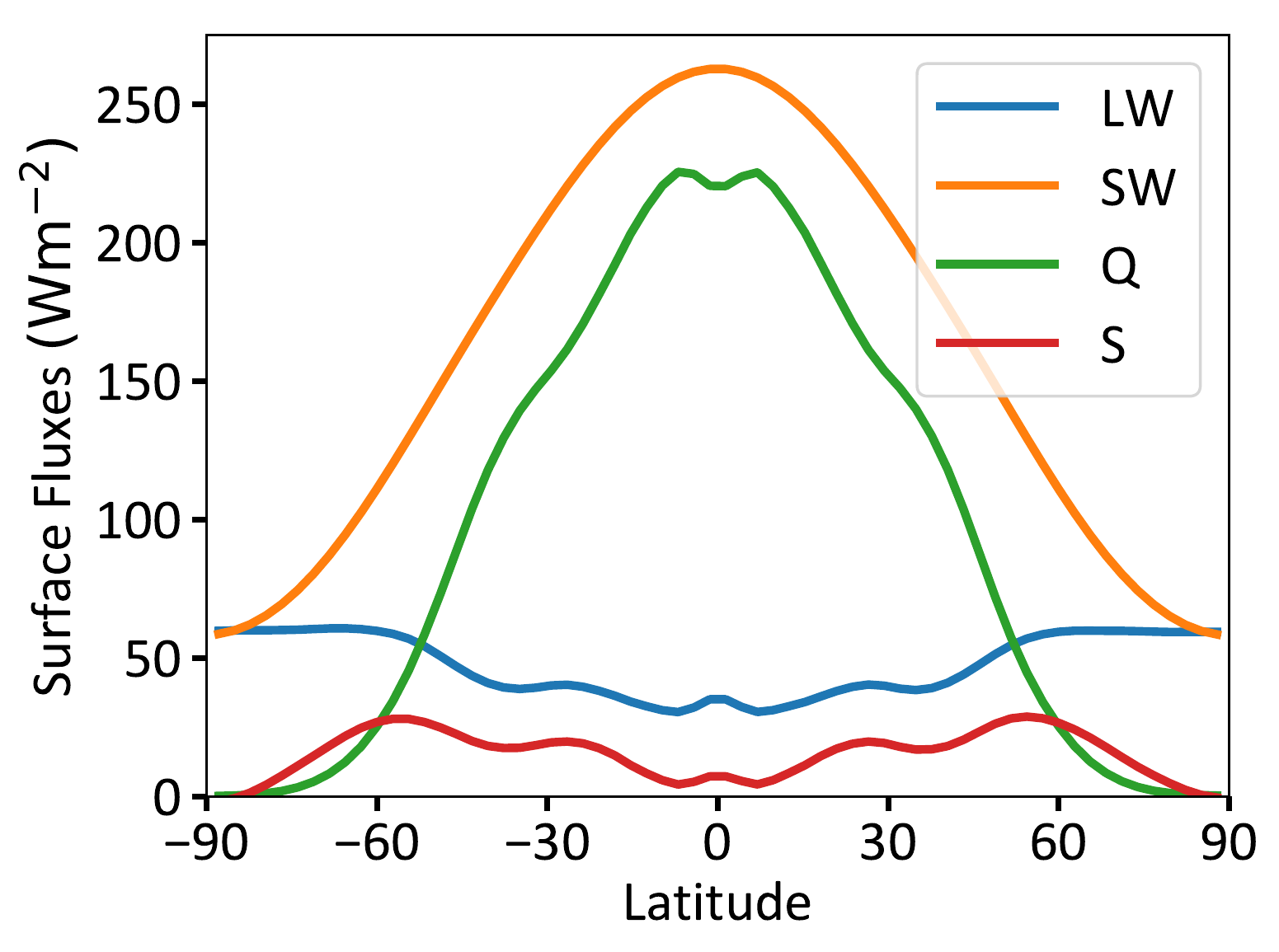}%
  \includegraphics[width=0.49\textwidth]{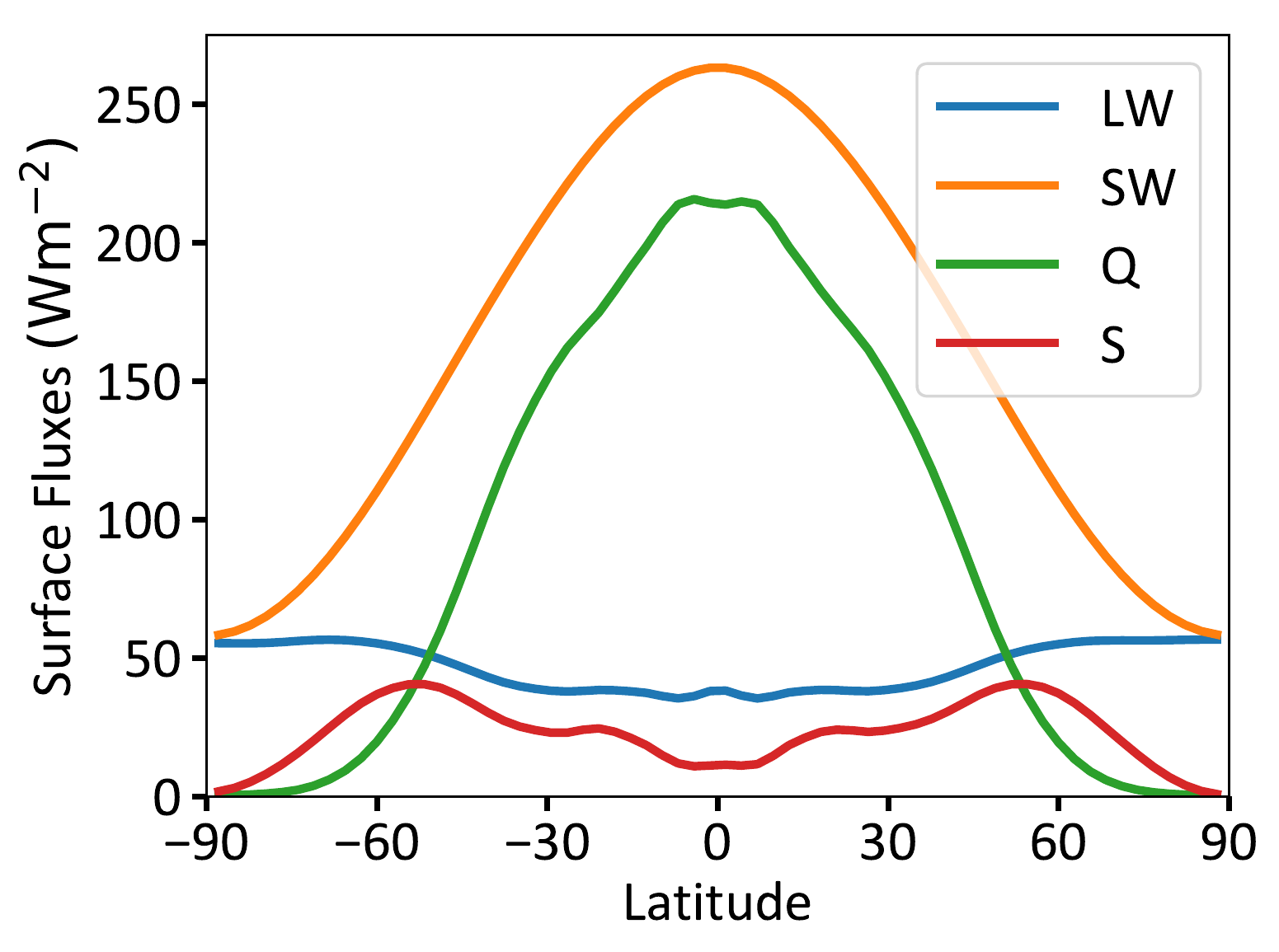}
\\
  \includegraphics[width=0.49\textwidth]{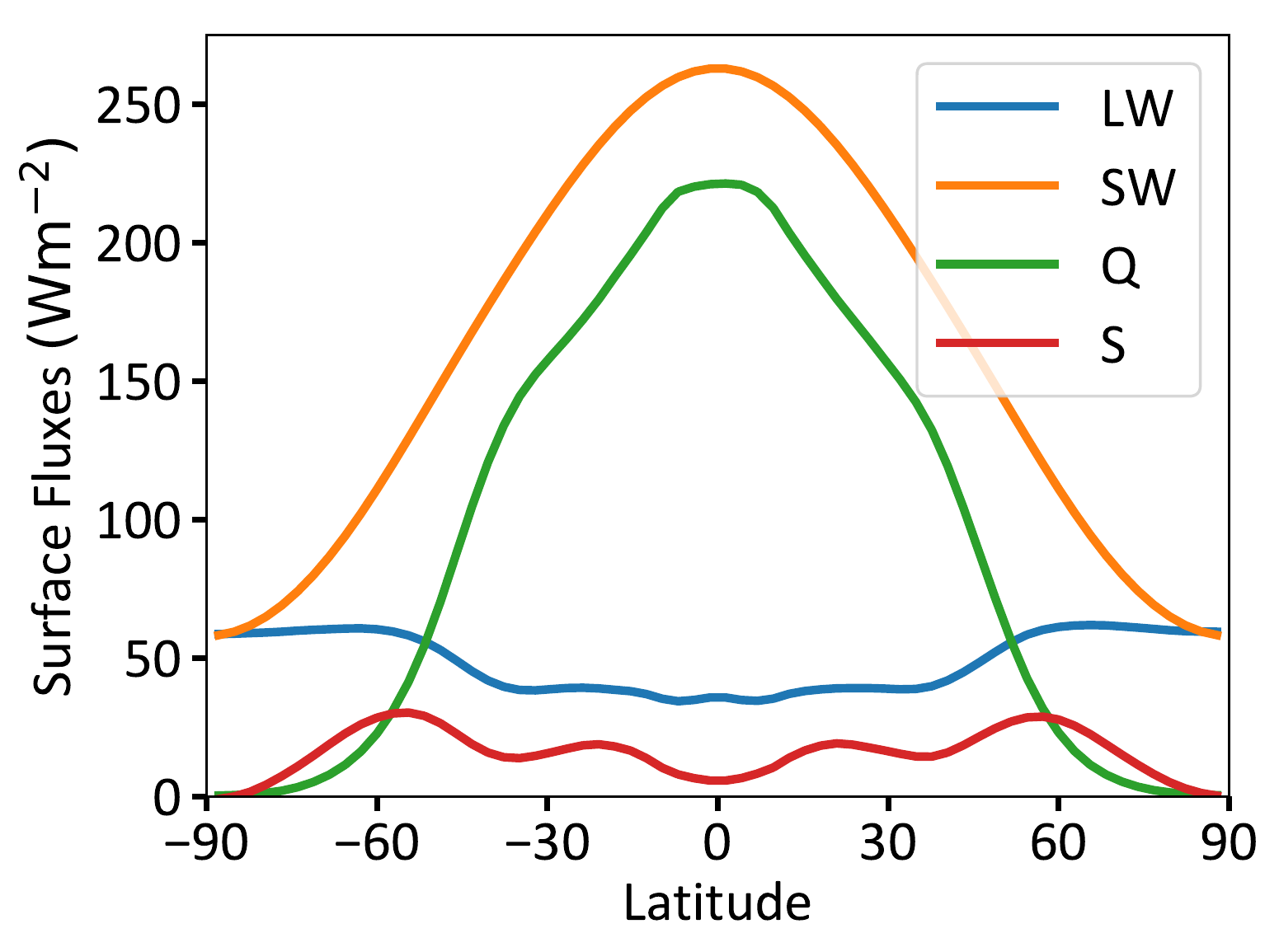}
\caption{\label{fig:frierson_surface_budget} \panel{a} the terms in the surface-temperature equation for a simulation with Earth-gravity. \panel{b} the same for twice-Earth gravity. \textsf{LW} and \textsf{SW} are the long and short-wave fluxes, \textsf{Q} is the latent heat flux and \textsf{S} the sensible heat flux. \panel{c} is the same as \panel{b} but from the twice-Earth gravity case in the scaled-surface-flux experiments described in section \protect\ref{sec:scaling_experiment}.
}
\end{figure}
}

The surface fluxes in the model are generalizations of this and are described by bulk-aerodynamic formulae with coefficients determined by Monin-Obukhov scaling, namely
\begin{subequations}
    \label{eq:surface_fluxes}
\begin{align}
    \sens & =  \rho_{\textrm{atm}} C_p C(z, z_{\textrm{sens}}) |\mathbf{V_a}| \left(\theta_{\textrm{surf}}- \theta_{\textrm{atm}}\right) \\
    \latent & = \rho_{\textrm{atm}} C(z, z_{\textrm{moist}}) |\mathbf{V_a}| \left(q_{\textrm{surf}}- q_{\textrm{atm}}\right) \\
    \tau & =  \rho_{\textrm{atm}} C(z/z_t) |\mathbf{V_a}| \mathbf{V_a},
\end{align}
\end{subequations}
where $\sens$ is the sensible heat flux out of the surface, $\latent$ is the latent heat flux out of the surface, and $\tau$ is the stress exerted by the surface on the atmosphere. In these formulae $\rho_{\textrm{atm}}$ is the atmospheric density at the lowest model level, $\mathbf{V_a}$ is the horizontal wind velocity on the lowest model level, $C_p$ is the heat capacity of dry air, $\theta_{\textrm{surf}}$ and $\theta_{\textrm{atm}}$ are the surface and lowest-model-level potential temperatures, respectively, $q_{\textrm{surf}}$ and $q_{\textrm{atm}}$ are the surface and lowest-model-level specific humidities, respectively, where $q_{\textrm{surf}}$ is the saturated specific humidity at the temperature of the surface . $C(z, z_{\textrm{rough}})$ is a function of the stability of the boundary layer as calculated by Monin-Obukhov similarity theory, where $z$ is the height on the lowest model level and $z_{\textrm{rough}}$ is the roughness length appropriate for each quantity. (In the simulations we take $z_{\textrm{sens}} = z_{\textrm{moist}} = z_{\textrm{stress}} = z_{\textrm{rough}} = 3.21\times 10^{-5} \textrm{m}$, and scale these values like $1/\alpha$ with changing gravity). 

The acceleration provided by $\tau$ in the momentum equations is $\rho^{-1} \partial \tau / \partial z$. Because $\tau$ scales like $\alpha$ in (\ref{eq:surface_fluxes}c), and the factors associated with $\rho^{-1} \partial / \partial z$ cancel out, the momentum tendency scales like $\alpha$.   This is equivalent to having an eddy diffusivity $\kappa$, that scales like $\sim 1/\alpha$, as before, and this holds for velocity, temperature and specific humidity.

The difference between surface latent heat fluxes and temperature fluxes lies solely in the different ways that $T$ and $q$ scale under a change in gravity, not in their effective eddy diffusivities.  Specifically, the latent heat flux given by (\ref{eq:surface_fluxes}b) remains invariant under a change in gravity, whereas the sensible heat flux scales like $\alpha$, because $\rho$ scales like $\alpha$.   These changes in surface fluxes affect the surface (mixed-layer) temperature which obeys an equation of the form
\begin{equation}
    C_{\textrm{surf}} \frac{\partial T_s}{\partial t} = SW - LW -\sens - \latent, 
    \label{eq:surface_temperature}
\end{equation}
where $C_{\textrm{surf}}$ is the mixed-layer's heat capacity, $T_s$ is the surface temperature, $SW$ is the net SW flux into the surface, $LW$ is the net long-wave cooling of the surface.
The gravitational acceleration is not explicitly present on the left-hand side, which will go to $0$ in a steady-state regardless, or in $SW$ or $LW$, and changes in $g$ will have no effect at lowest order on these terms.  As discussed above, $\latent$ has no dependence on $\alpha$, but $\sens$ increases like $\alpha$. Thus, under a increase in gravity we expect that latent heat fluxes to play a relatively smaller role in the heat balance of the surface layer.

To see these various effects, \reff{fig:frierson_surface_budget} shows the time-averaged terms on the RHS of \refeqq{eq:surface_temperature} in a case with normal-Earth gravity in panel \panel{a} and twice-Earth gravity in panel \panel{b}.

The short-wave heating of the surface is the same in both cases, as our radiation scheme has a  fixed solar absorption, but there are notable changes in the other flux components. The sensible heat fluxes, $\sens$,  increase everywhere because of the changes in the atmospheric density, as is predicted by the simple scaling. In the polar regions, where the latent cooling of the surface is small, the increase in $\sens$ necessitates a decrease in the long-wave cooling of the surface in order to maintain a balance (in the surface heat budget) with the short-wave heating. This decrease in long-wave cooling is achieved by a decrease in both the atmospheric and surface temperatures. This is consistent with the polar cooling seen in \reff{fig:frierson1}.

\begin{figure}
\centering
\subfloat[]{
\includegraphics[width=0.5\textwidth]{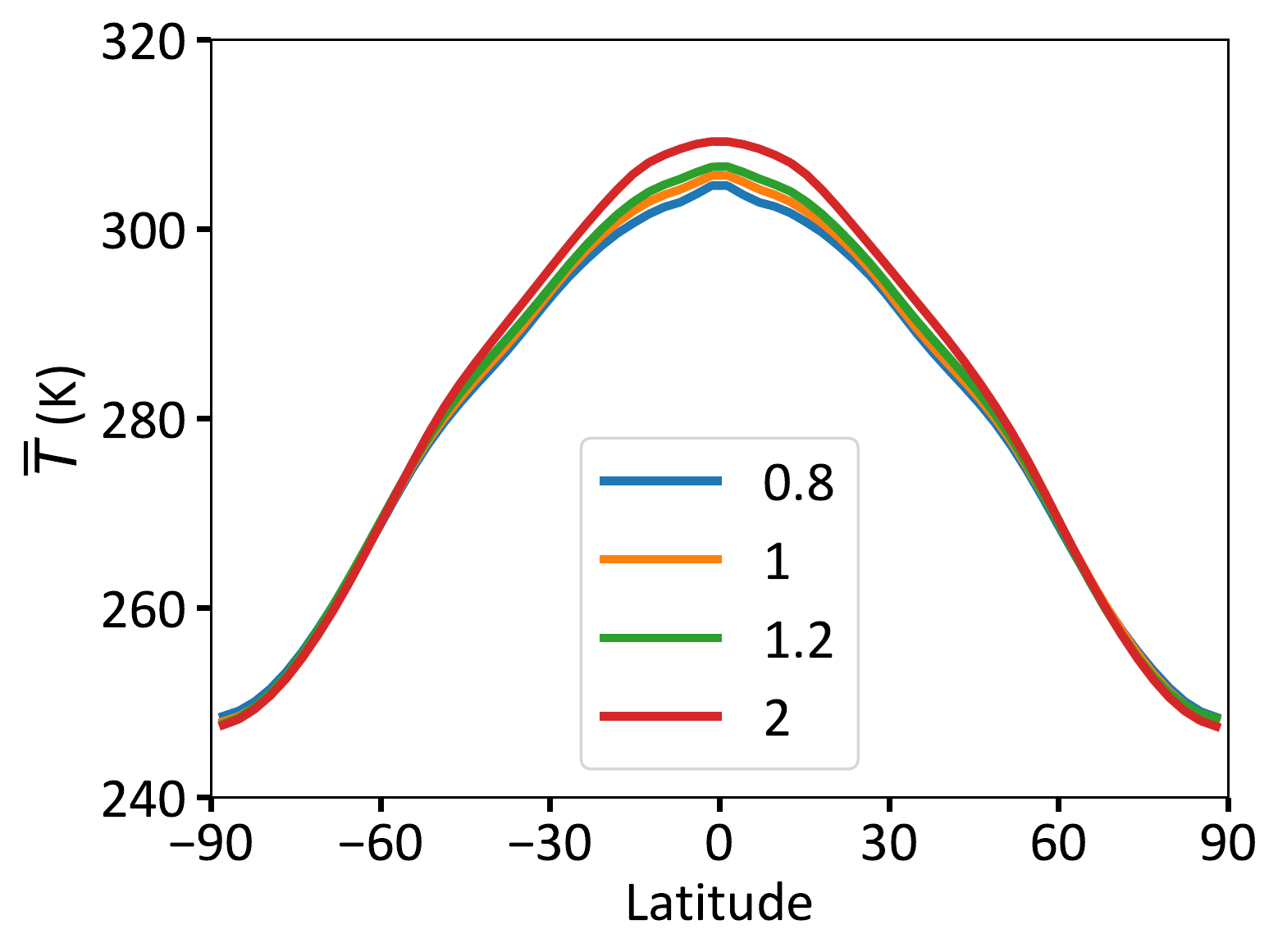}
}
\subfloat[]{
  \includegraphics[width=0.5\textwidth]{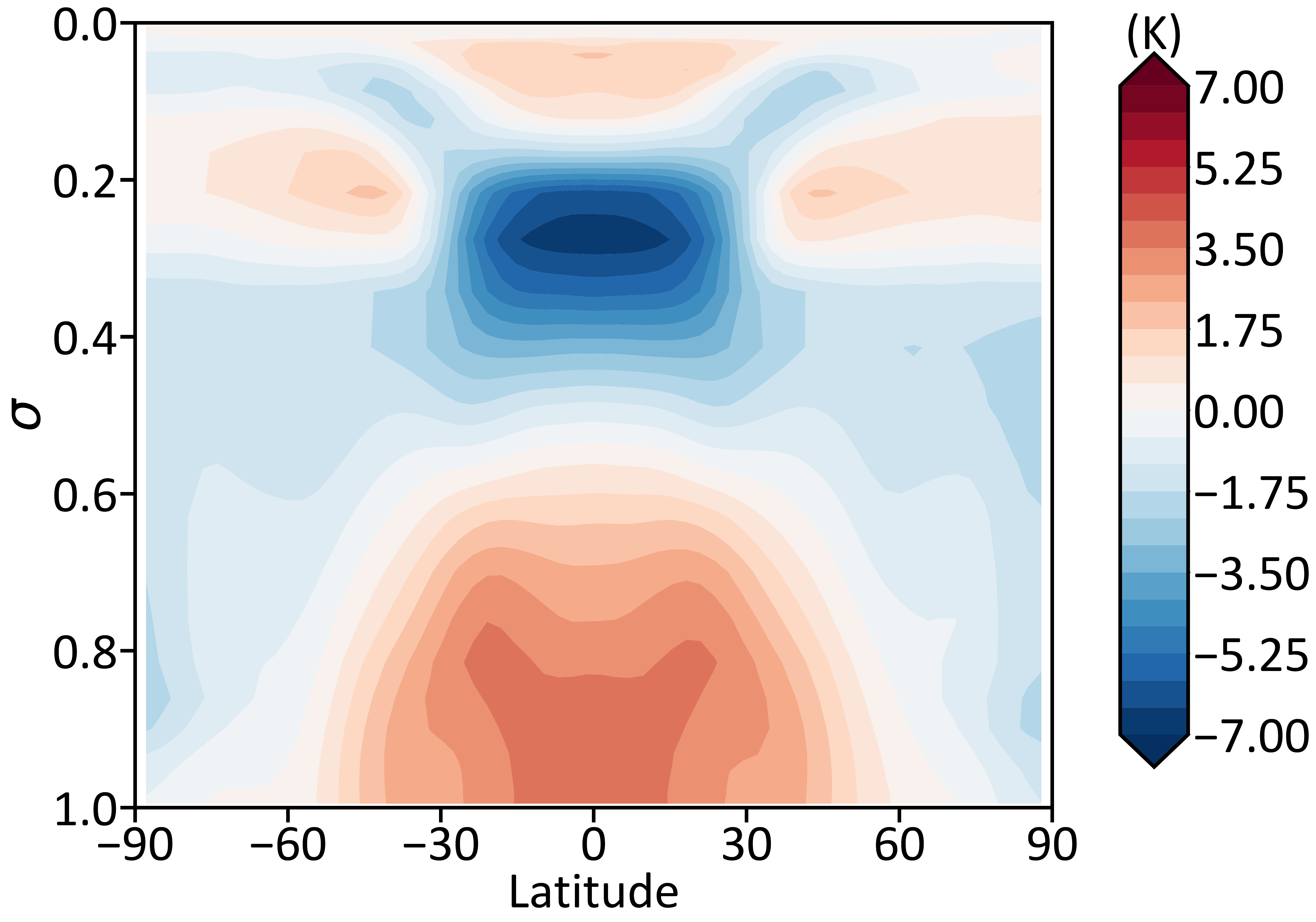}
}
\caption{\label{fig:frierson_scale_heat_and_mom} Panel \panel{a} shows the zonal-mean surface temperature against latitude in aquaplanet run with optical depths like \citet{Frierson_etal06}, but also with scaled $\sens$ and $\tau$ so that they do not scale with gravity. Panel \panel{b} shows the zonal-mean atmospheric temperatures in a twice-Earth-gravity run minus the same in a normal-Earth-gravity run, plotted as a function in $\sigma$ coordinates.}
\end{figure}

In the tropics, the latent heat fluxes decrease slightly with increased gravity. The scaling result is that they stay constant, and the decrease arises because $|\mathbf{V_a}|$ decreases with increased gravity. This is because of the increased surface stress $\tau$ due to the increased atmospheric density, which leads to weaker near-surface winds. The decrease in tropical latent heat fluxes is partly offset by the increase in tropical sensible heat fluxes, but the sum of the two decreases when gravity is doubled, necessitating an increase in long-wave cooling, which is provided by way of surface and the lower tropical atmosphere warming. It is clear that the changes in the surface energy budget are consistent with the temperature changes seen in \reff{fig:frierson1}.

Note that in the experiments described above, the contribution of $C(z, z_{\textrm{rough}})$ does not change significantly with gravity. This is partly due to our scaling of $z_\textrm{rough}$ like $1/\alpha$, but is also a reflection that the stability of the boundary layer does not change significantly with gravity. 

\subsection{A scaling experiment} 
\label{sec:scaling_experiment}

In order to isolate the influence of the changes in surface fluxes with changed gravity, we conduct an experiment where $\rho_{\textrm{atm}}$ in the $\sens$ and $\tau$ formulas is divided by $\alpha$, so that $\sens$ and $\tau$ no longer scale proportionally with gravity. We leave the $\rho_{\textrm{atm}}$ in $\latent$ as it is, so that none of the three fluxes then scale with gravity. The results of this experiment are shown in \reff{fig:frierson_scale_heat_and_mom}.

Comparing the scaled-flux results in \reff{fig:frierson_scale_heat_and_mom} with the unscaled-flux results in \reff{fig:frierson1}, it is clear that the polar cooling apparent in the unscaled experiment is no longer present in the scaled experiment.  Analysis of the surface energy budget in this case, shown for the twice-Earth-gravity case in \reff{fig:frierson_surface_budget} \panel{c}, is consistent with the polar cooling in the unscaled case being caused by a change in the sensible heat fluxes.  The tropical latent heat fluxes do change by a small amount in the scaled experiments, but by a smaller amount than in the unscaled experiments. This is consistent with part of this change being due to the changed surface winds in the unscaled experiment. It is also clear, particularly from a comparison of \reff{fig:frierson1}\panel{c} with \reff{fig:frierson_scale_heat_and_mom}\panel{b}  that the changes in tropical temperatures are not caused by the scaling of surface fluxes with gravity.

\section{Lapse-rate Changes}
\label{sec:tropics}

To investigate the changes in tropical temperatures associated with changing the gravitational constant, we turn to analysing changes in the tropical lapse rates. On Earth, the tropical lapse rate remains close to the moist-adiabatic lapse rate, which is calculated based on the condensation of ascending saturated parcels. In climate change projections, it is well-known that a significant component of the tropical temperature changes are caused by changes in the tropical lapse rate and the tropopause height \citep[see e.g.][]{Vallis_etal15}. Warming (associated with increased greenhouse gases), leads to a higher moisture content and a decreased magnitude of the moist-adiabatic lapse rate, and so more warning in the upper tropospheric regions than at the surface.  In addition, because the atmosphere remains in radiative balance with the incoming short-wave radiation, the temperature of the tropopause  (which in a grey atmosphere is directly related to the emission temperature) stays constant. A corollary of this is that the tropopause height increases under with global warming.  Although the argument is only exact for grey radiation, similar effects are seen in GCMs with full radiation schemes. 

The same process, but in reverse, is operating in our experiments with increased gravity that include water-vapour optical depth feedback, as seen in \reff{fig:bog1}. Here, increased gravity decreases surface moisture, and so decreases long-wave optical depth, leading to a colder surface. The decreased surface moisture increases the magnitude of the moist-adiabatic lapse rate and the upper troposphere cools more than the surface. The tropopause height decreases under increased gravity, to maintain a constant outgoing longwave radiation. 

The experiments with no water-vapour optical depth feedback do not get a surface cooling under increased gravity (\reff{fig:frierson1} and \reff{fig:frierson_scale_heat_and_mom}). However, the specific humidity does decrease considerably (\reff{fig:frierson1}\panel{d}). Consequently the magnitude of the moist adiabatic lapse rate increases, leading to a cooler upper troposphere and a lower tropopause.

To quantify these notions we construct simplified tropical temperature profiles using the following assumptions. 
\begin{itemize}
    \item The stratosphere is optically thin and in radiative balance such that it has a constant temperature equal to the emission temperature.
    \item Radiative transfer is grey in the infra-red, with a surface optical depth of $\tau_s$. 
    \item The lapse rate, $\Gamma_r$,  is a constant in height.
\end{itemize}    
The tropopause height can then be calculated according to the following equation,  from \citet{Vallis_etal15}. 
\begin{equation}
    H_T = \frac{1}{16\Gamma_r}\left(C \ttrop + \sqrt{C^2\ttrop^2 + 32 \Gamma_r \tau_s H_a\ttrop}\right)
    \label{eq:tropopause_height}
\end{equation}
Here $H_T$ is the height of the tropopause, $\Gamma_r$ is a representative lapse rate, $C=2\log 2 \approx 1.38$, $H_a$ is the scale height of the atmospheric absorber, and $\ttrop$ is the tropopause temperature, which can be approximated via the incoming solar radiation assuming that the stratosphere is optically thin.
The tropospheric temperature profile, $T(z)$, can then be calculated using 
\begin{equation}
    T(z) = \ttrop - \Gamma_r(q_r) (H_T-z),
    \label{eq:constant_lapse_rate_temp}
\end{equation}    
where $z$ is the height above the surface.

\begin{figure}
\centering
\subfloat[]{
\includegraphics[width=0.5\textwidth]{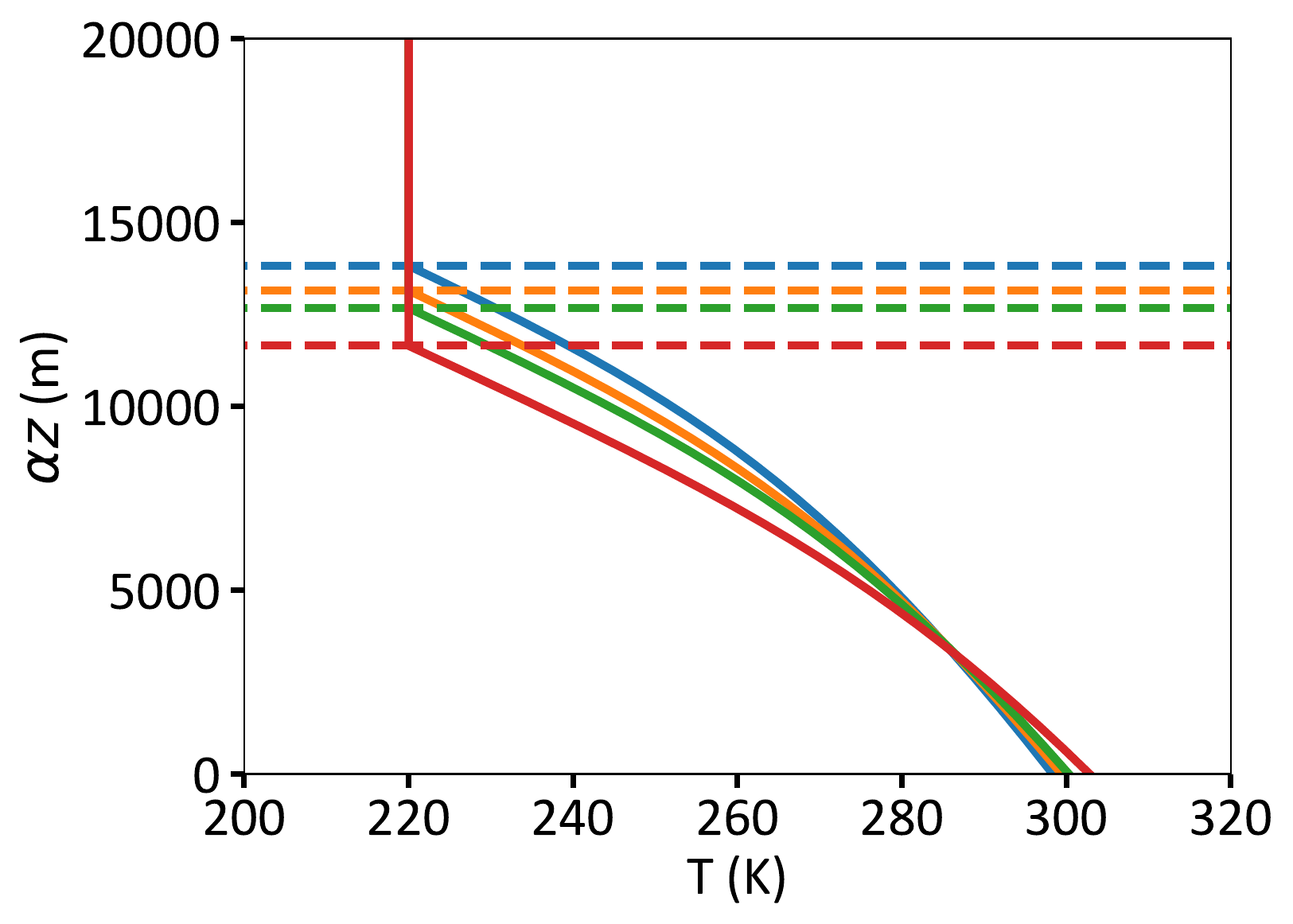}
}
\subfloat[]{
  \includegraphics[width=0.5\textwidth]{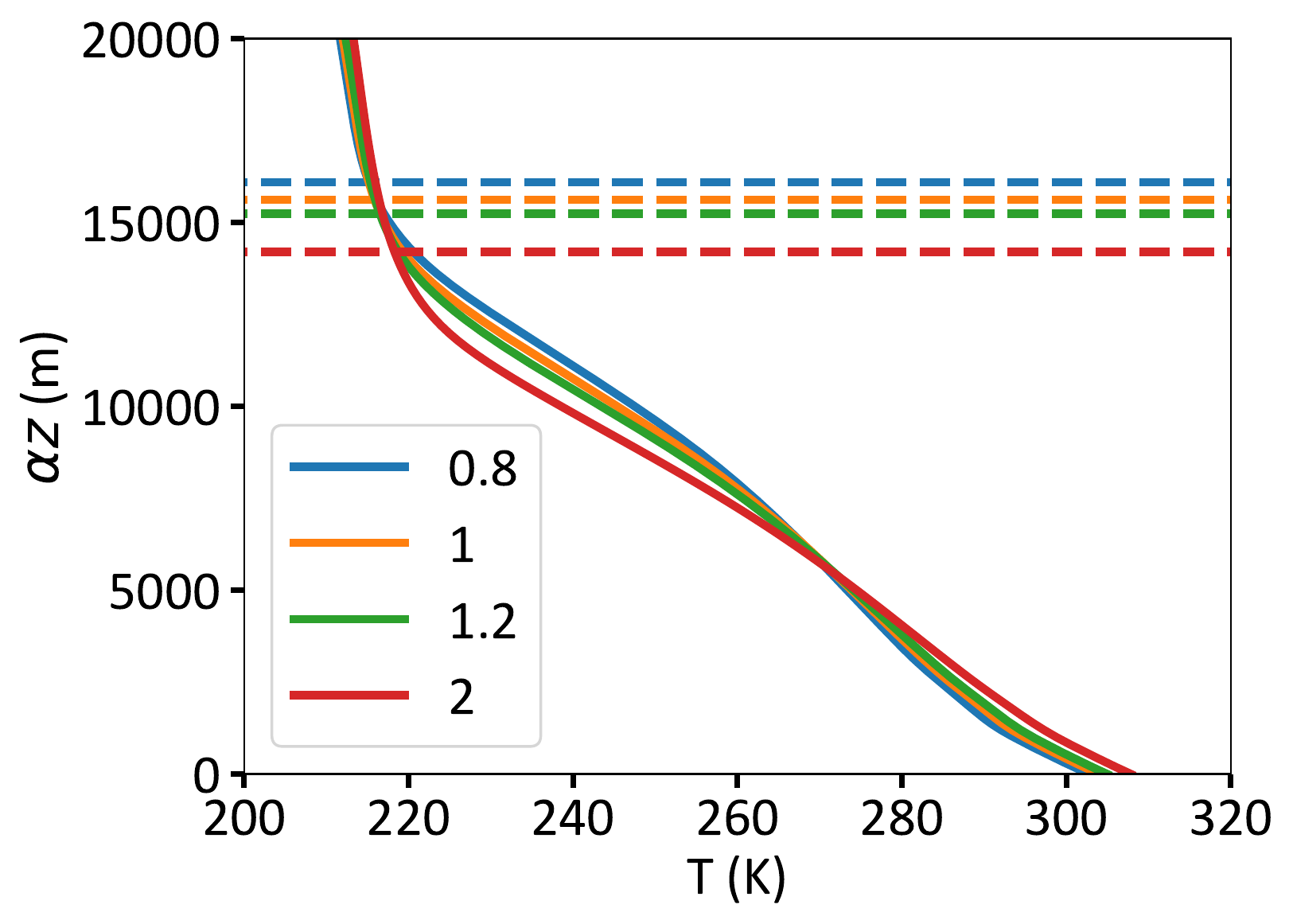}
}
\caption{\label{fig:frierson_recon_temp_profiles} Panel \panel{a} shows representative temperature profiles constructed using  \protect\refeqq{eq:varying_lapse_rate_temp} and related assumptions for Earth gravity and twice-Earth gravity. The horizontal dashed lines are the tropopause heights for the two gravity values, and the solid lines are the temperature profiles. The $z/\alpha$ axis has been scaled by $1/\alpha$ so that the profiles can be compared on the same height scale. Panel \panel{b} shows temperature profiles from the surface-flux-scaled experiments, thereby isolating the lapse-rate effects. These profiles have bee averaged between 10S and 10N, and have their tropopause heights calculated from the 2K/km threshold definition.}
\end{figure}

A slight extension to this formalism in useful, in which we continue to use \refeqq{eq:tropopause_height} for $H_T$, but the vertical profile for temperature is constructed with a vertically-varying lapse rate, $\Gamma_s$, which can be taken to be the saturated adiabatic lapse rate. The temperature profile is then given by
\begin{equation}
    T(z) = \ttrop - \Gamma_s(q_s(z),T(z)) (H_T-z),
    \label{eq:varying_lapse_rate_temp}
\end{equation}    
where $q_s(z)$ is the saturation specific humidity at height $z$. In our calculations of $H_T$ we take $\Gamma_r=\Gamma_s$ calculated using $T=255K$ and $q_s$ calculated at a representative pressure of $\sigma=0.3$, making $\Gamma_r$ close to $6\alpha K/km$, which is a representative value for Earth. Using \refeqq{eq:varying_lapse_rate_temp} and (\ref{eq:tropopause_height}), we construct representative temperature profiles for the range of $\alpha$ values used in our experiments, which are shown in \reff{fig:frierson_recon_temp_profiles}\panel{a}. These artificial profiles show that, under an increase in gravity, $\Gamma_s$ increases, the tropopause height drops, and surface temperatures warm, with temperatures in the upper troposphere falling.

In \reff{fig:frierson_recon_temp_profiles}\panel{b} we show time and latitude averaged vertical temperature profiles from our grey-radiation experiments with scaled surface fluxes, as shown in \reff{fig:frierson_scale_heat_and_mom}. These are qualitatively similar to those in panel \panel{a}, verifying that effects included in our artificial profiles, i.e. changes in the tropical lapse rate and tropopause height, are sufficient to explain the temperature changes seen in \reff{fig:frierson1} and \reff{fig:frierson_scale_heat_and_mom}. 

In contrast to the above, let us also consider artificial profiles in regions far from saturation, where $\Gamma\approx\Gamma_d=g/c_p$, so that $T(z) = \ttrop - \Gamma_d (H_T-z)$.  This dry adiabatic lapse rate increases like $\alpha$, so at a given value of $z/\alpha$ the atmospheric temperatures would be the same independent of changes in gravity.  (This result is actually demanded by the fact that the dry equations are invariant with respect to changes in $\alpha$, with $z\to z/\alpha$.)
Therefore the lapse-rate effects seen in \reff{fig:frierson_recon_temp_profiles} should not be present in regions that are far from saturation, hence why the lapse-rate changes are only apparent in the tropics in \reff{fig:frierson1} and \reff{fig:frierson_scale_heat_and_mom}. 

\section{Responses with realistic radiative transfer}
\label{sec:socrates}

In the above sections we isolated various effects using idealized models. We now use a  more realistic model to  explore their combined effect, and in particular we use the \Socrates radiative transfer code \citep{Edwards1996, Manners_etal17}. \Socrates is a highly flexible radiative transfer code that has been used extensively in operational UK Met Office models and in the study of exoplanetary atmospheres \citep[e.g.][]{Amundsen2016}. Here we use 
\Socrates with 12 long-wave bands and 21 short-wave bands, and run it without ozone absorption in the stratosphere for ease of comparison across different gravity values. We also run without a seasonal-cycle, instead forcing the model with the incoming short-wave profile used in the grey-radiation experiments.

\afterpage{
\begin{figure}[H]
\centering
\includegraphics[width=0.49\textwidth]{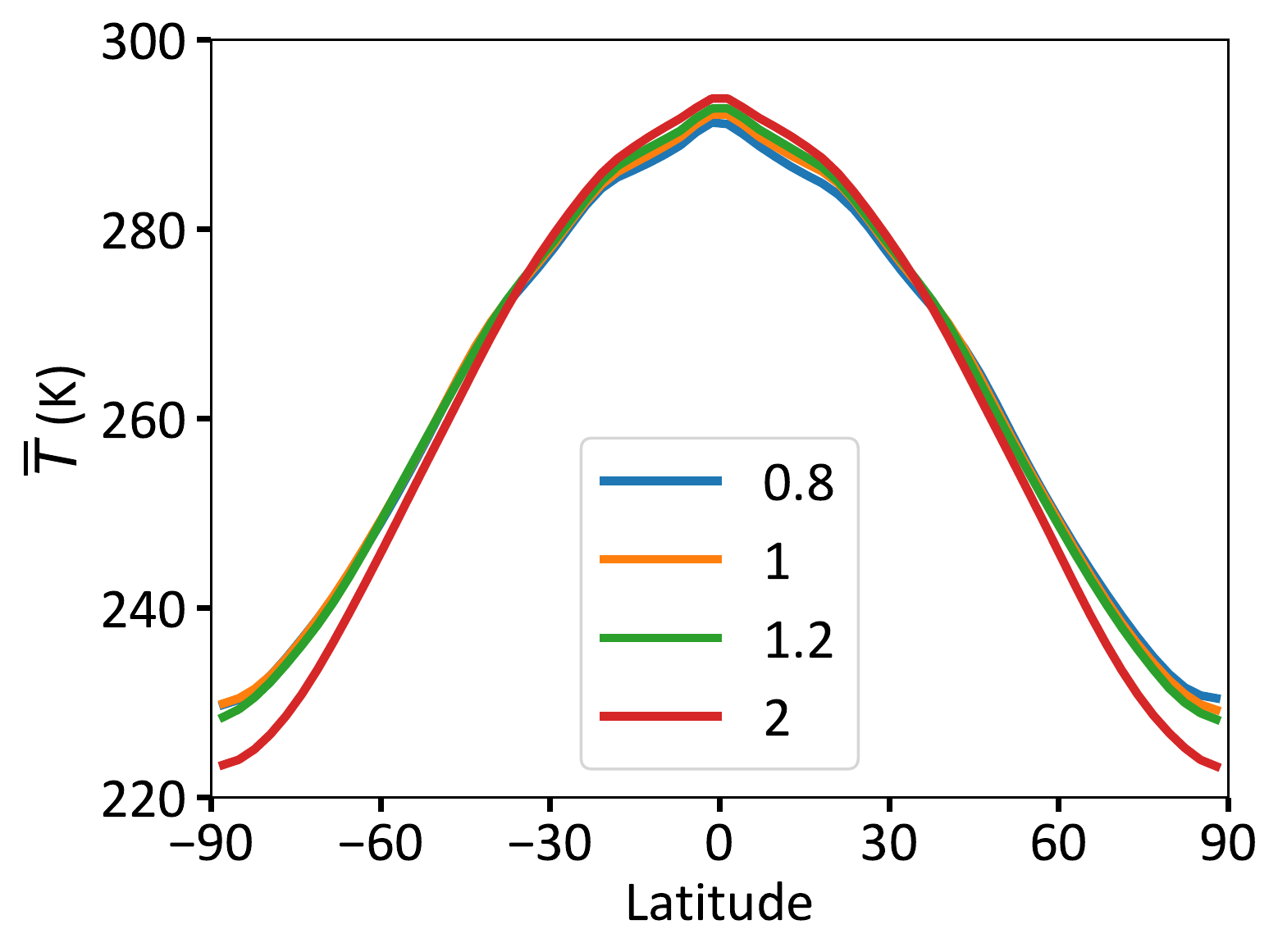}v%
  \includegraphics[width=0.49\textwidth]{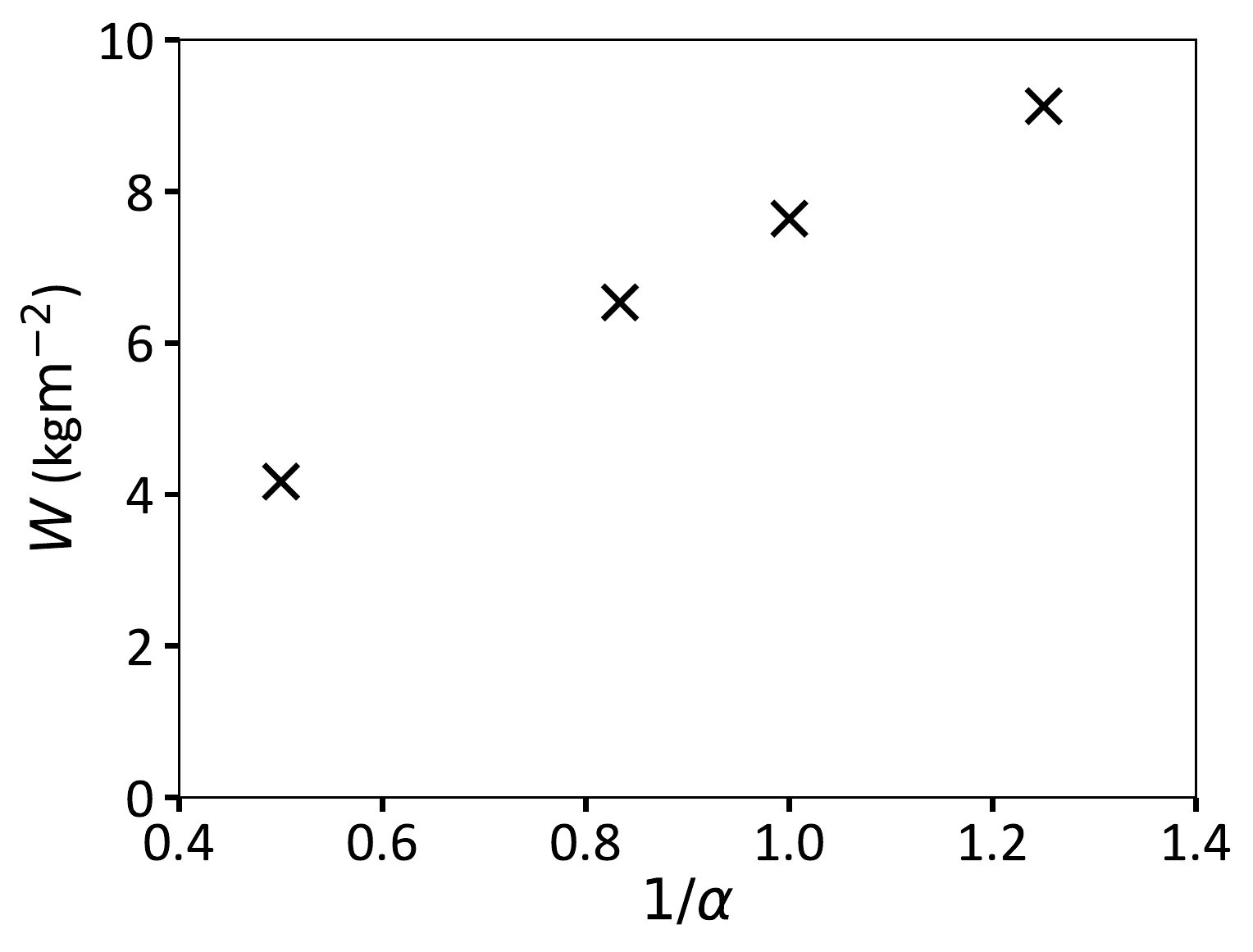}%
  \\[0.1cm]
  \includegraphics[width=0.49\textwidth]{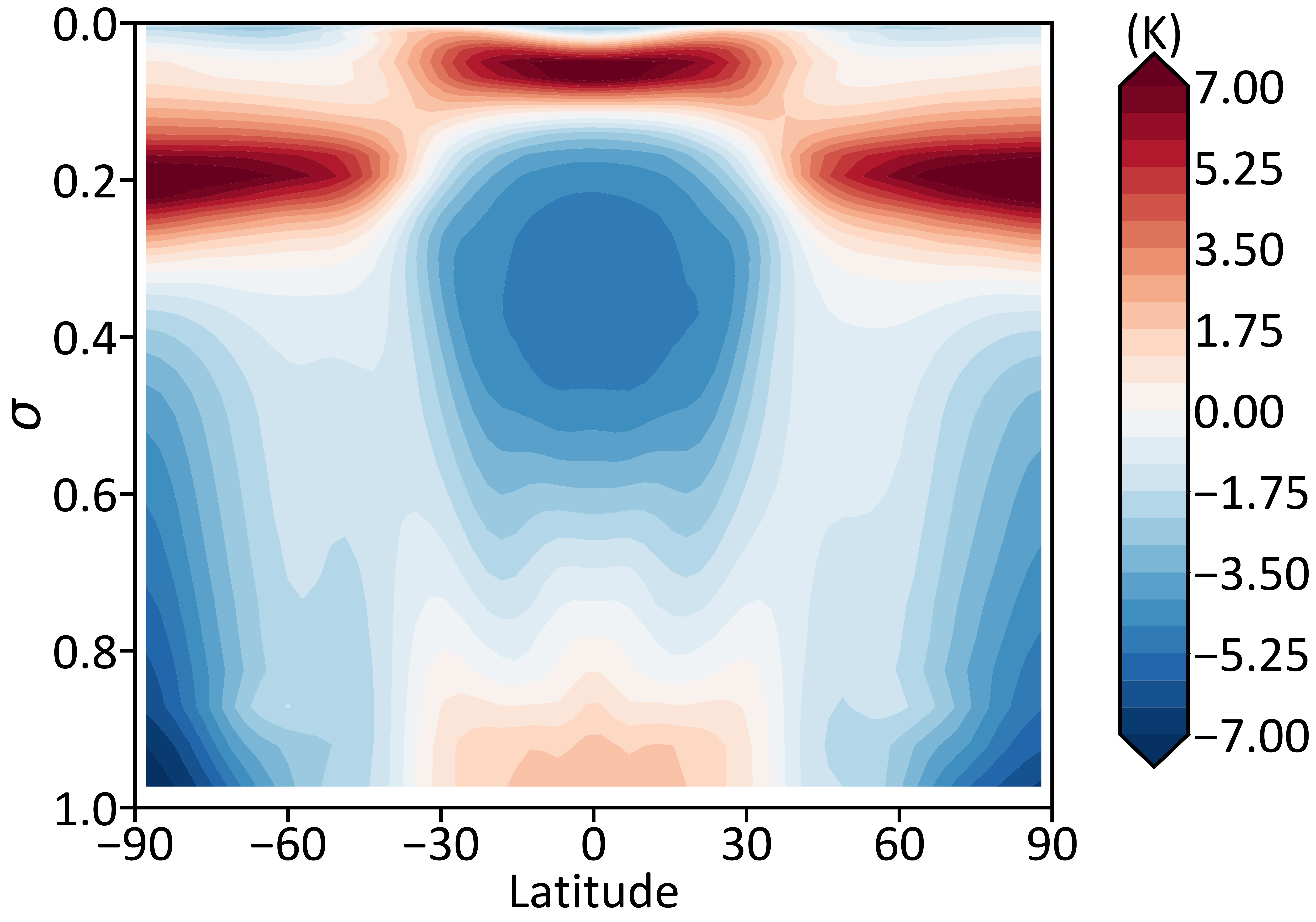}
\caption{\label{fig:socrates} Panel \panel{a} The zonal-mean surface temperature against latitude in aquaplanet run with the \Socrates radiation scheme. Panel \panel{b} The time and area-averaged value of $W$ plotted against $1/\alpha$. Panel \panel{c} The zonal-mean atmospheric temperatures in a twice-Earth-gravity run minus the same in a normal-Earth-gravity run, plotted as a function in $\sigma$ coordinates.}
\end{figure}
}

The zonal-mean surface temperatures in these experiments are shown in \reff{fig:socrates}\panel{a}. The Earth-like climate is colder with \Socrates than it was with the grey radiation schemes of \citet{Frierson_etal06} shown in \reff{fig:frierson1} and \reff{fig:frierson_scale_heat_and_mom}, and is somewhat similar to the temperatures in \reff{fig:bog1}. This is also reflected in the total water-vapour amounts shown in panel \panel{b}, which are  lower with \Socrates than it was with the grey radiation schemes. The difference in temperature in the Earth-like cases is likely because of the increased absorption of short-wave radiation in \Socrates compared with the grey schemes, and the inclusion of the well-known spectral window for long-wave cooling in \Socrates.

Despite the mean-state differences, a number of familiar features are apparent in the atmospheric temperature response to a doubling of gravity shown in panel \reff{fig:socrates}\panel{c}. The cooling of the surface outside of the tropics is consistent with a decrease in long-wave optical depth due to decreased column moisture. In the tropical regions we see warming at the surface and in the lower troposphere with cooling aloft, consistent with an increase in the saturated adaibatic  lapse rate changes, but with an additional cooling  due to the long-wave optical depth decrease. In contrast with \reff{fig:bog1}, the decrease in long-wave optical depth is not enough to offset the surface heating from the lapse rate changes, meaning that a realistic combination of these effects is not quite the same as an inverse of a climate-change response.

The equator to pole heat transport decreases in these experiments (not shown) consistent with this increase in equator-to-pole temperature gradient. This result was not found in the grey radiation runs. The lack of transport change with grey radiation may well be a special case for the grey radiation prescription and parameters that are used, consistent with the contrasting transport changes found with grey radiation in \citet{Frierson_etal07} and \citet{Schneider_etal10}.

The latent and sensible heat fluxes changes with \Socrates are broadly similar to those shown in \reff{fig:frierson_surface_budget}, with a diminished role for latent heat fluxes under increased gravity.  One notable difference is that the sensible heat fluxes are (with \Socrates) negative in the polar regions, because the atmosphere is warmer than the surface. As a result, the increase in the magnitude of the sensible heat fluxes, due to the $\alpha$ scaling described above, necessitates an increase in surface long-wave cooling in the polar regions (whereas previously it gave rise to a decrease). 

Finally, we note that an increase in gravity gives rise to an increase in pressure broadening of the spectral lines in the radiative transfer, an effect only included in our runs with \Socrates. This broadening is related to the absolute atmospheric pressure i.e. $p$ not $p/p_s$, so that a higher gravity gives rise to higher pressure and more broadening. This turns out to be a small effect compared with the reduction of water vapour condensible, and we do not describe the results. However, it may play a more important role in a dry atmosphere.

\section{Conclusions}
\label{sec:conclusions}

In this paper we have investigated the response of terrestrial atmospheres to a change in the gravitational acceleration at the planet's surface. The full Navier--Stokes momentum equations in spherical geometry do have a dependence on gravity, but these changes are usually small in a terrestrial atmosphere, consistent with the (normally very good) approximations used to derive the primitive equations. The primitive equations remain invariant under a transformation where gravity is changed by a factor $\alpha$ if  the vertical co-ordinate $z$ is scaled by a factor $1/\alpha$. Any changes found due to changes in gravity must then arise from thermodynamical and radiative aspects of the planet's atmosphere, and their interaction with the dynamics, rather than the dynamics alone. 

The effects of a change in gravity on an Earth-like atmosphere arise from two main phenomena:
\begin{enumerate}
    \item A change in the total column water vapour under gravity, arising from a change in the atmospheric scale height combined with the scaling invariance of the temperature field. Thus, in a higher gravity planet the atmosphere has a smaller vertical extent and less total water vapour. Since water vapour is a potent greenhouse gas this effect leads to an overall cooling of the atmosphere.
    \item A change in the specific humidity, at least in a dilute atmosphere in which the condensible is a small fraction of the total atmosphere. In such an atmosphere $q \approx \eps\vp/p$, and since $p$ scales with gravity while $\vp$ does not, an increase of gravity leads to a general reduction in specific humidity. Thus, an increase in gravity leads to a reduction in the effects of condensation, with the following main effects:
\begin{enumerate}
    \item Changes in specific humidity lead to changes in the saturated adiabatic lapse rate, which is the dominant factor determining lapse rate in the tropics. An increase in gravity leads to warming near the surface and cooling aloft. This effect is very robust across all experiments and parameters.
    \item Changes in surface sensible heat flux, surface stress, and latent heat fluxes from the surface, with the first two scaling like $\alpha$ but the third not. 
    A reduction in $q$ at higher gravity thus leads to a reduction in the relative importance of the latent heat flux compared to sensible heat flux. The effects of this are rather complicated, and lead to different amounts of long-wave cooling and polar cooling under higher gravity, differing quantitatively across experiments with different radiation schemes.
    \item Changes in the relative components of the meridional energy flux. A reduction in $q$ at higher gravity leads to a smaller meridional latent heat flux, but in many experiments this is compensated by an increase in the sensible heat flux. We do not ascribe a universality to this result. 
\end{enumerate}
\end{enumerate} 
In addition, changes in the pressure broadening of spectral lines due to changes in atmospheric pressure have a small effect. In our experiments this effect is much smaller than the change in greenhouse effect due to changes in the amount of condensible, but in a dry atmosphere the effect would be dominant. 

The balance between the above effects will determine the overall response, and that balance is determined by the properties of the atmosphere and condensible. In this paper we have focussed on an Earth-like planet, but a condensible with a smaller latent heat content than water, but a larger effect on the long-wave optical depth, would make the radiative effects more important than the condensation effects.   

In a non-dilute atmosphere --- that is, one in which the condensible is not a minor constituent --- the effects would be different again since the approximation leading to \eqref{spec.1} is no longer valid and the relative amount of the condensible would not necessarily change with gravity.  However, assuming that the amount of condensible is determined primarily by the planetary temperature, a reduction in scale height of the atmosphere with increased gravity would still lead to a smaller total amount of condensible, and (if the condensible is a greenhouse gas) to a cooler planet (and then still less condensible).  Evidently, the properties of any condensible species are key in setting the atmospheric temperature structure and its circulation for any given planet or exoplanet.

\subsection*{Acknowledgements}
The authors wish to thank Bob Beare for useful discussions, and Mark Hammond and James Manners for their help with setting up \Socrates. The authors declare no conflict of interest. Documentation and code required to run the Isca model framework can be found at \url{www.exeter.ac.uk/isca} and \url{www.github.com/ExeClim/Isca}.

\bibliography{allrefs}
\end{document}

%% file: mypreamble.tex
\newcounter{regequation}

\newcommand{\newcm}[2]{\newcommand*{#1}{\ensuremath{#2}}}

\newcommand{\newc}{\newcommand*}

\newenvironment{notes}{  \begin{small} \begin{sffamily}}%
 {\end{sffamily} \end{small}}

 {\end{enumerate} \end{notes}}

  {\setcounter{equation}{\value{regequation}}
    \end{shaded}\end{minipage}\end{center}\end{table} }

  {\setcounter{equation}{\value{regequation}}
    \end{shaded}\end{minipage}\end{center}\end{table} }

 {\setcounter{equation}{\value{regequation}}
   \end{boxedminipage}\end{center}\end{table} }

\newcommand{\x}{$x$\xspace} \newcommand{\y}{$y$\xspace}  \newcommand{\z}{$z$\xspace}

\newcommand{\probref}[1]{\thechapter.\ref{#1}}
\newcommand{\letrn}[2]{\lettrine[lines=3,lhang=0]{{#1}}{#2}}
\newcommand{\figref}[1]{fig.\ \ref{#1}}
\newcommand{\Figref}[1]{Fig.\ \ref{#1}}
\newcommand{\secref}[1]{Section \ref{#1}}

\newcommand{\dbint}{\iint}

\newcommand{\curlz}{\text{curl}_z}
\renewcommand{\div}[1][]{\nabla_{#1}\cdot}

\newcommand{\omb}{\bm {\omega}}
\newcommand{\Omb}{\bm {\Omega}}

\newcommand{\taub}{\bm{\tau}}

\newcommand{\s}{\, \text{s}} 
\newcommand{\m}{\, \text{m}}
\newcommand{\cm}{\, \text{cm}\xspace}
\newcommand{\km}{\, \text{km}\xspace}
\newcommand{\kg}{\, \text{kg}} \newcm{\ps}{\, \text{s}^{-1}}
\newcommand{\Kv}{\, \text{K}}
\newcommand{\qg}{quasi-geostrophic\xspace}
\newcommand{\pg}{planetary geostrophic\xspace}
\newcommand{\BV}{Brunt-V\"ais\"al\"a\xspace}
\newcommand{\crn}{\nonumber  \\}
\newcommand{\qqquad}{\qquad \qquad}
\newcommand{\cdel}{\cdot \nabla}
\newcommand{\Ro}{\ensuremath{\mathit{Ro}}\xspace}
\newcommand{\Fr}{\ensuremath{\mathit{Fr}}\xspace}
\newcommand{\Bu}{\ensuremath{\mathit{Bu}}\xspace}
\newcommand{\Ri}{\ensuremath{\mathit{Ri}}\xspace}
\newcommand{\Rot}{\ensuremath{\mathit{Ro_\mathrm{t}}}\xspace}

\newcommand{\bpp}[2]{\bfrac{\partial #1}{\partial #2}}
\newcommand{\bfrac}[2]{\Bigg( \frac{#1}{#2} \Bigg) }
\newcommand{\bpps}[3][]{\Bigg({\partial #2 \over \partial #3}\Bigg)_{\! #1}}
\newcommand{\Eta}{\mathcal{E}}
\newcommand{\K}{\mathcal{K}}
\newcommand{\hhat}{\ensuremath{\hat h}}
\newcommand{\fhat}{\ensuremath{\hat f}}

\newcommand{\psisp}[1]{\psi^{(#1)}}
\newcommand{\psihat}{\hat \psi}
\newcommand{\qhat}{\hat q}
\newcommand{\psibar}{\overline \psi}
\newcommand{\phibar}{\overline \phi}
\newcommand{\thetabar}{\overline \theta}
\newcommand{\Psihat}{\hat \Psi}
\newcommand{\Psibar}{\overline \Psi}
\renewcommand{\hbar}{\overline h}

\newcommand{\qbar}{\overline q}
\newcommand{\ubar}{\overline u}
\newcommand{\vbar}{\overline v}

\newcommand{\bbar}{\overline b}
\newcommand{\bybar}{\overline \by}

\newcommand{\zetabar}{\overline \zeta}
\newcommand{\betahat}{\ensuremath{\hat \beta}}
\newcommand{\kbeta}{\ensuremath{k_\beta}\xspace}

\newcommand{\etahat}{\hat \eta}
\newcommand{\ubhat}{{\hat{\bm{u}}}}
\newcommand{\xhat}{{\hat x}}
\newcommand{\yhat}{{\hat y}}
\newcommand{\zhat}{{\hat z}}
\newcommand{\uhat}{\hat u}
\newcommand{\vhat}{\hat v}
\newcommand{\that}{{\hat t}}
\newcommand{\khat}{\hat k}
\newcommand{\bhat}{\hat b}
\newcommand{\fbhat}{\hat{\bm{f}}}
\newcommand{\phihat}{\hat \phi}
\newcommand{\byhat}{\hat \by}
\newcommand{\ug}{\ensuremath{u_g}}
\newcommand{\vg}{\ensuremath{v_g}}
\newcommand{\Lr}{\ensuremath{L_R}}
\newcommand{\lr}{\ensuremath{l_r}}
\newcommand{\lrr}{\ensuremath{\lambda_r}}
\newcommand{\vrt}{\vartheta}
\newcommand{\cross}{\bm{\times}}
\newcommand{\half}{\tfrac 12}

\newcommand{\dt}{\, \mathrm{d}t}
\newcommand{\dx}{\, \mathrm{d}x}
\newcommand{\dy}{\, \mathrm{d}y}
\newcommand{\dz}{\, \mathrm{d}z}
\newcommand{\dk}{\, \mathrm{d}k}
\newcommand{\dr}{\, \mathrm{d}r}
\newcommand{\dkb}{\, \mathrm{d}\bm{k}}
\newcommand{\dxb}{\, \mathrm{d}\xb}

\newcommand{\dAb}{\, \mathrm{d} \bm{A}}
\newcommand{\Dth}{{\D_h \over \D t}}
\newcommand{\Dto}{{\D_0 \over \D \that}}
\newcommand{\Fb}{\bm{F} }
\bmdefine\Fb{F}

\newc\Ab{\ensuremath{\bm{A}}}
\newc\Jb{\ensuremath{\bm{J}}}
\newc\Bb{\ensuremath{\bm{B}}}
\newc\xb{{\bm{x}}}
\bmdefine{\ab}{a}
\bmdefine{\Xb}{X}
\bmdefine{\Yb}{Y}
\def\rb{{\bm{r}}}
\bmdefine{\sb}{s}
\bmdefine{\Db}{D}

\bmdefine{\Vb}{V}
\bmdefine{\Wb}{W}
\def\lb{{\bm l} }
\newcommand{\fb}{{\ensuremath{\bm {f}}}}
\newcommand{\fbo}{{\ensuremath{\bm {f_0}}}}
\def\vb{{\ensuremath{\bm {v}}}\xspace}
\def\ub{{\ensuremath{\bm {u}}}\xspace}
\def\wb{{\ensuremath{\bm {w}}}\xspace}

\newcommand{\psib}{\ensuremath{\bm \psi}\xspace}

\newcommand{\by}{\ensuremath{b}}
\newcommand{\Tb}{{\bm T} }
\newcommand{\lhs}{left-hand-side}
\newcommand{\fo}{\ensuremath{f_0}\xspace}
\def\DD#1{{\D#1 \over \D t}}

\newcommand{\MD}[1]{{\partial {#1} \over \partial t} + (\vb \cdot \del) #1}

\newcommand{\pp}[3][]{{\partial^{#1} #2 \over \partial #3^{#1}}}

\newcommand{\DDD}[1]{\D{#1} / \D t}
\newcommand{\ppa}[1]{{\partial \over \partial #1}}
\newcommand{\pt}{{\partial \over \partial t}}
\newcommand{\del}{\nabla }
\newcommand{\grad}{\nabla}
\newcommand{\delh}{\nabla_h}
\newcommand{\cp}{c_p}

\newcommand*{\eps}{\epsilon}

\renewcommand{\d}{{\mathrm{d}}}
\newcommand{\D}{{\mathrm{D}}}

\newcommand{\textmonth}{\ifcase\month \or January \or February\or
March\or April\or May \or June \or July\or August\or September\or October
\or November\or December\fi\xspace}

\DeclareMathAlphabet{\mathsfsl}{\encodingdefault}{hlst}{b}{sl}   

\newcommand{\putabcd}{\put(-10.5,8.8){\small \textit{\textsf{(a)}}}
\put(-5,8.8){\small \textit{\textsf{(b)}}}
\put(-10.5,4.2){\small \textit{\textsf{(c)}}}
\put(-5,4.2){\small \textit{\textsf{(d)}}}
}

\newcommand{\bfig}{\begin{figure}}
\newcommand{\efig}{\end{figure}}
\newcommand{\bcen}{\begin{center}}
\newcommand{\ecen}{\end{center}}
\newcommand{\beql}[1]{\begin{equation}\label{#1}}
\newcommand{\beqn}{\begin{eqnarray}}
\newcommand{\eeqn}{\end{eqnarray}}
\newcommand{\beq}{\begin{equation}}
\newcommand{\eeq}{\end{equation}}
\newcommand{\benu}{\begin{enumerate}}
\newcommand{\benui}{\begin{enumerate}[{(}i{)}]}
\newcommand{\benuI}{\begin{enumerate}[I]}
\newcommand{\eenu}{\end{enumerate}}
\newcommand{\bite}{\begin{itemize}}
\newcommand{\eite}{\end{itemize}}
\newcommand{\bdes}{\begin{description}}
\newcommand{\edes}{\end{description}}
\newcommand{\bsub}{\begin{subequations}}
\newcommand{\esub}{\end{subequations}}
